\documentclass[prl,aps,twocolumn,showpacs,superscriptaddress]{revtex4-2}
\usepackage[USenglish]{babel}
\usepackage{textcomp} 
\usepackage{amsmath,dsfont}
\usepackage{amsmath, amsthm, amssymb}
\usepackage{xr}
\usepackage[bbgreekl]{mathbbol}
\usepackage{graphicx}
\usepackage{dcolumn}
\usepackage{bm}
\usepackage{xcolor}
\usepackage{comment}
\usepackage{hyperref}

\begin{document}


\title{Mechanics and wrinkling patterns of pressurized bent tubes}
\author{César L. Pastrana}
\affiliation{Physics of Complex Biosystems, Technical University of Munich, 85748 Garching, Germany}
\author{Luyi Qiu}
\affiliation{John A. Paulson School of Engineering and Applied Sciences, Harvard University, Cambridge, Massachusetts 02138, USA}
\author{John W. Hutchinson}
\affiliation{John A. Paulson School of Engineering and Applied Sciences, Harvard University, Cambridge, Massachusetts 02138, USA}
\author{Ariel Amir}
\affiliation{John A. Paulson School of Engineering and Applied Sciences, Harvard University, Cambridge, Massachusetts 02138, USA}
\affiliation{Department of Physics of Complex Systems, Weizmann Institute of Science, 7610001 Rehovot, Israel}
\author{Ulrich Gerland}
\affiliation{Physics of Complex Biosystems, Technical University of Munich, 85748 Garching, Germany}


\begin{abstract}
Take a drinking straw and bend it from its ends. After sufficient bending, the tube buckles forming a kink, where the curvature is localized in a very small area. This instability, known generally as the Brazier effect, is inherent to thin-walled cylindrical shells, which are particularly ubiquitous in living systems, such as rod-shaped bacteria. However, tubular biological structures are often pressurized, and the knowledge of the mechanical response upon bending in this scenario is limited. In this work, we use a computational model to study the mechanical response and the deformations as a result of bending pressurized tubes. In addition, we employ tension-field theory to describe the mechanical behaviour before and after the wrinkling transition. Furthermore, we investigate the development and evolution of wrinkle patterns beyond the instability, showing different wrinkled configurations. We discover the existence of a multi-wavelength mode following the purely sinusoidal wrinkles and anticipating the kinked configuration of the tube.  
\end{abstract}

\maketitle

\begin{figure}[b!]
    \centering
    \includegraphics[width=0.49\textwidth]{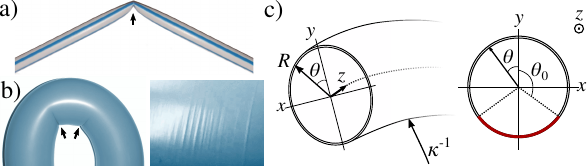}
    \caption{Buckling of thin tubes under bending loads. 
    a) Buckled drinking straw. 
    b) \emph{left.} Buckling in a pressurized party balloon. Arrows highlight the buckled regions. \emph{right.} Wrinkles observed at the bottommost side of a pressurized and bent party balloon, anticipating the buckling shown in the left panel.
    d) Geometric variables. The wrinkled area (red) spans the domain $\theta>\theta_0$ and  $\theta<-\theta_0$, with $\theta \in [-\pi,\pi$).}
    \label{fig:intro_notation}
\end{figure}

\emph{Introduction}.---Exerting bending loads on a thin-walled tube elastically deforms the tube by curving it. If bending increases, the curvature reaches a critical value where an instability takes place, in which the curvature is localized on small regions of the surface \citep{calladines_brazier_chapter}. This phenomenon can be observed in the sudden formation of a kink by bending a drinking-straw (Fig.~\ref{fig:intro_notation}a). 
\par 
Brazier studied the buckling of tubular shells subjected to bending, finding that the relation between torque and curvature is not monotonic \citep{brazier_main}. A first type of instability is known as Brazier effect and is expected to occur at the curvature giving a maximum torque. Nonetheless, a second type of instability, known as the wrinkling or birfucation instability, can occur at a torque (and curvature) lower than that expected by Brazier's, which results in wave-like structures immediately before the formation of the final buckled shape \citep{karamanos_bending}. Depending on the geometric and mechanical properties of the shell, different shapes are observed beyond the instability, ranging from kinked configurations to wrinkled or triangular (Yoshimura) patterns \citep{ashkan_tR_patterns}.
\par
The formation of kinks or wrinkles in tubes subjected to bending loads is not restricted to the macroscale~\citep{mahadevan_inext_wrinkle_analysis}. At the nanoscale, wrinkled and/or kinked structures have been observed in carbon nanotubes and microtubules \citep{buckling_carbon_nanotube_1, buckling_carbon_nanotube_2, mahadevan_microtubules}. It is therefore expected that instabilities should also occur at intermediate scales.  Rod-shaped bacteria are an interesting example of thin-walled tubes at the microscale. Yet, bacterial cells present a remarkable distinction compared to the previous examples: the differential concentration of osmolites between the interior and the exterior of the bacteria results in significant internal (turgor) pressures \citep{Beveridge_afm_Youngs}, such that they are deformed by pressure akin to a party balloon (Fig.~\ref{fig:intro_notation}b). Notably, cylindrical thin shells subjected to pressures are common in biological systems, e.g., blood vessels or the absorbent hairs in the roots of plants \citep{blood_vessels_review, root_hairs_diameter}. This situation is in stark contrast with that typically found in engineering problems, e.g. pipes, where the large stiffness of the material prevents significant deformations \citep{yuan2014_basic_mechs_pressure_tubes, kyriakides2007}. 
\par 
Here, we investigate the response of thin-walled pressurized tubes under bending loads. We develop a computational model permitting us to extract the mechanical response, and to contrast it with recent analytical formulas \citep{bending_luyis}. Furthermore, using a theoretical framework based on tension-field theory, we extend the characterization of the torque-curvature relation beyond the wrinkling point. We find that the results observed in the computational model are well described by our theoretical framework, including the expected curvature at wrinkling. 
Since the behavior of the system beyond the instability is challenging to describe analytically, we computationally characterize the properties of the wrinkles, their dependency on pressure, and how the wrinkled state evolves with additional bending post-buckling. We show that two different wrinkled states can be formed beyond the instability and prior to the final kinked configuration. In particular, we observe a sinusoidal regime with low amplitude wrinkles, followed by a multi-wavelength pattern including large amplitude wrinkles anticipating the kinked configuration. 
\par
\textit{Model}.---We study tubes with unpressurized radius $R_0$ and length $L_0 \gg R_0$ described by a triangulated mesh with two flat lids that close the volume of the tube. The total energy of the system is given by $\mathcal{E} = \mathcal{E}_\textrm{int} + \mathcal{E}_\textrm{ext}$. The internal (mechanical) energy of the pressurized shell is given by $\mathcal{E}_\textrm{int} = \mathcal{E}_\textrm{s} + \mathcal{E}_\textrm{b} - pV$ where $p$ is the pressure and $V$ the enclosed volume \citep{pressure_induced_paper}. The stretching and bending energies, $\mathcal{E}_\textrm{s}$ and $\mathcal{E}_\textrm{b}$, are described by harmonic potentials and penalize in-plane and out-of-plane deformations of the surface (Secs. I-III of the Supplemental Material, \citep{sm_ref}). This model maps to a continuum material \citep{seoung_nelson_flexible_membranes}. Taking bacteria as inspiration, we use the characteristic 3D Young modulus $E$ and thickness $t$ of Gram negative cell walls \citep{beveridge_afm}. The potential $\mathcal{E}_\textrm{ext}$ penalizes the misalignment between the normal vectors of the lids with respect to defined target vectors. This term is introduced to induce the overall bending of the tube, acting as virtual hands. We determine the resulting configurations at a given pressure by minimizing $\mathcal{E}$ at progressively increasing degrees of bending.
\par
\textit{Results}.---The buckling of a tube in the absence of pressure has been widely studied during the last century \citep{calladines_brazier_chapter, brazier_main}. Therefore, we used the unpressurized condition as a proof-of-principle of the ability of the computational model to produce well-established results. We find quantitative agreement with the response described in the literature for the torque-curvature ($M-\kappa$) curves as well as for the geometry/ovalization of the cross-section (with radii $R_x$ and $R_y$), see Sec. IV of the Supplemental Material \citep{sm_ref}. Furthermore, at the theoretically expected wrinkling curvature the shell collapses in a kinked configuration resembling that of bent drinking straws (Fig.~\ref{fig:intro_notation}a). 
\par
\begin{figure}[t!]
    \centering
    \includegraphics[width=0.49\textwidth]{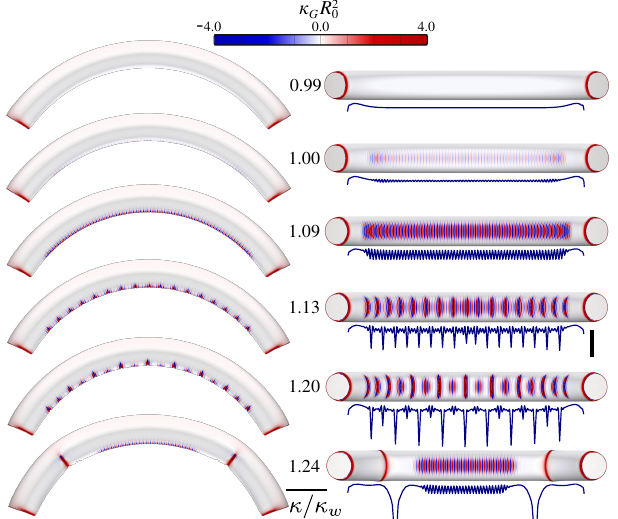}
    \caption{Wrinkle patterns during bending. 
    Resulting configurations after the wrinkling transition for $\hat{p} \equiv pR/(Et) = 0.026$ at progressively increasing curvatures $\kappa/\kappa_w$ from top to bottom, where $\kappa_w$ is the curvature at the onset of wrinkling.  The surface is colored according to Gaussian curvature, $\kappa_G$ \citep{meyers_diff_geo_mesh}. 
    Blue continuous lines show the radius at the bottom of the tube $R_y$ along the arc length of the tube. The scale bar indicates $R_y/R_0 = 0.05$.
    The pictures at $\kappa/\kappa_w = 1.13$ and $\kappa/\kappa_w = 1.20$ are representative examples of the system in the multi-wavelength state (MWS) at the low and high wavelength substates, respectively.}
    \label{fig:wrinkles_general}
\end{figure}
Having verified that the numerical model is consistent with established theoretical predictions, we investigate the much less studied response of tubes in a pressurized state. We study a regime of pressures leading to radial expansions in the range 1-10\%. We find that after sufficient bending, the tube acquires a wrinkled configuration (Fig.~\ref{fig:wrinkles_general}, top rows).  The onset of the wrinkling instability is marked by the sudden appearance of periodic structures at the bottom of the tube (Supplemental Material and Movies S1-S3, \citep{sm_ref}). Contrary to what is observed at zero pressure, the wrinkled state persists with additional bending, developing a complex wrinkling pattern before collapsing into a kinked configuration where the curvature is highly localized (Figs. \ref{fig:wrinkles_general}, bottom row). That is, the wrinkling transition is not concomitant with the kinking instability, as occurs at zero pressure. Furthermore, kinked tubes again develop sinusoidal wrinkles in the space between kinks, and this is observed for every pressure \footnote{Since the mesh model does not include long range interactions, the surface can intersect with itself near the kinks. We therefore focus our analysis on the regime before the appearance of kinks.}.
\par
\begin{figure*}[t!]
    \centering
    \includegraphics[width=0.99\textwidth]{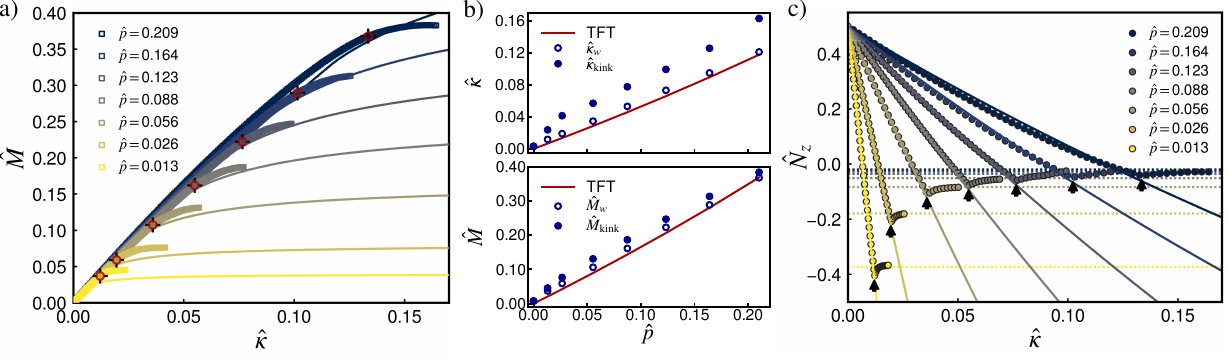}
    \caption{Mechanics of pressurized bent tubes. 
    a) Torque and curvature. The observed buckling position for each pressure is indicated with red bullets. Theoretical predictions from the tension-field theory (TFT) are shown with continuous lines (Eqs. \ref{eq:torque_pre_w} and \ref{eq:torque_post_w}). 
    b) Curvature (\emph{top}) and torque (\emph{bottom}) at the onset of wrinkling (open symbols) and at the kink instability (filled symbols). The continuous lines are the predictions for $\hat{\kappa}_w$ and $\hat{M}_w$ from TFT ($\hat{\mathcal{K}}_w = \hat{p}/2$ and  $\hat{M}_w = \pi\hat{p}/2$). 
    c) Stress resultants. The dimensionless resultant stress is defined as ${\hat{N} \equiv N/(pR)}$. Longitudinal stress resultants $\hat{N}_{z}$ on the inner-most region of the tube as a function of curvature $\hat{\kappa}$. Continuous lines are the predictions from Eq.~\ref{eq:Nz}. The critical stresses $N_c$ are indicated with dotted lines (Eq. \ref{eq:sigma_c}). The observed onset of wrinkling for each pressure is marked with a black arrow. 
    }
    \label{fig:mechanics_press_tubes}
\end{figure*}
\par 
Next, we characterized the mechanical response to bending. From the torque-curvature curves we find that for low bending, $M$ grows linearly with $\kappa$ (Fig.~\ref{fig:mechanics_press_tubes}a). The response continues until the tube develop wrinkles, from which point further bending results in sublinear increase of $M$ with $\kappa$, eventually plateauing near the kinking instability. We determine the curvature and torque at the onset of wrinkling, $\kappa_w$ and $M_w$ as well as that at the onset of the kinking instability, $\kappa_\mathrm{kink}$ and $M_\mathrm{kink}$ (Fig.~\ref{fig:mechanics_press_tubes}b). We find that the wrinkled state exists for a larger range of $\kappa$ beyond the wrinkling transition for high $p$. As a result of the saturating behavior of the torque with $\kappa$, the difference in torque between $M_w$ and $M_\mathrm{kink}$ is small and shows little dependency on $p$.
\par 
Recently, the mechanics of pressurized bent tubes has been studied analytically \citep{bending_luyis}. Here we further the understanding by including the response after the onset of wrinkling. To understand the mechanical response including the saturating post-wrinkling torque, we develop an analytical model using tension-field theory (TFT, \citep{tension_field_theory_steigmann}). The underlying assumption of TFT is that wrinkles accommodate extra area, relieving large negative stresses and minimizing the in-plane deformation energy.
On the unwrinkled region of the tube we assume linear relations between the resultant membrane stress and the strains. On the wrinkled region of the tube, we follow a TFT framework where we assume that the formation of wrinkles suppresses further increase of compressive stress with bending. 
Thus, the lateral stress $N_{z}$ remains constant at a critical compressive stress and the wrinkled region expands along the circumferential direction upon bending. 
The critical resultant membrane stress $N_c$ at the onset of wrinkling in an unpressurized tube is given by
\begin{equation}
    \label{eq:sigma_c}
    N_c = \frac{2}{R_{\theta}} \sqrt{BY},
\end{equation}
where $R_\theta$ is the circumferential radius of curvature at the position of maximal compression \citep{seide_weigarten_wavelength_bent_tube}.
Though $R_\theta$ can depend on $\kappa$ due to cross-section ovalization, our numerical results show that, at the range of $p$ studied, cross-section deformations are greatly suppressed by the internal pressure \citep{sm_ref}. Thus, we assume that the cross-section remains circular.
Considering linear elasticity, the resultant stress on the inner-most part of the tube ($\theta=\pi$, Fig.\ref{fig:intro_notation}c) before wrinkling is
\begin{equation}
    N_z = \frac{pR}{2} - EtR\mathcal{K}, 
    \label{eq:Nz}
\end{equation}
where $\mathcal{K}$ is the curvature of the medial axis, ~\mbox{$\mathcal{K} \equiv \kappa/(1 - \kappa R)$}. We calculate the stresses from the simulation and compare them to our predictions in Fig.~\ref{fig:mechanics_press_tubes}c. We observe a linear decrease of $N_{z}$ with $\kappa$, with values well-described by Eq.~\ref{eq:Nz}.  $N_{z}$ remains (mostly) constant with additional bending once the system wrinkles and the compressive stresses in the wrinkled configuration are in good agreement with the expected critical stress given by Eq.~\ref{eq:sigma_c}.
We can also calculate the $M$-$\kappa$ relationship within this model. For mathematical convenience, we approximate the critical wrinkling stress as zero, such that in the wrinkled region ($|\theta|>\theta_0$, where $\theta_0$ stands for the limiting extension of the wrinkles in the circumferential direction, Fig.~\ref{fig:intro_notation}c) the resultant membrane stress in the longitudinal direction equals zero, $N_z = 0$. We obtain a pressure-independent linear relation between torque and curvature prior to the onset of wrinkling (Fig.~\ref{fig:mechanics_press_tubes}a):
\begin{equation}
    \hat{M} = \pi\hat{\mathcal{K}}, 
    \label{eq:torque_pre_w}
\end{equation}
where we use non-dimensional variables defined as $\hat{\mathcal{K}} \equiv \mathcal{K} R$, $\hat{M} \equiv M/(EtR^2)$ and $\hat{p} \equiv pR/(Et)$. Beyond the wrinkling transition, we find
\begin{equation}
    \hat{M} =\left[\theta_0 - \frac{\sin(2\theta_0)}{2} \right]\hat{\mathcal{K}},
    \label{eq:torque_post_w}
\end{equation} 
where $\theta_0$ depends on $p$ and $\mathcal{K}$, see Sec. VI of the Supplemental Material \citep{sm_ref}. The torque-curvature relation for fixed $p$ determined within this model is plotted and compared to the simulation results in Fig.~\ref{fig:mechanics_press_tubes}a.
The computational results and the theoretical model are in good agreement, including the saturation of $M$ with additional bending past the wrinkling transition. In addition, Eq.~\ref{eq:Nz} can be used to calculate the expected curvature at the onset of wrinkling under the assumption that $N_z = 0$, leading to $\hat{\mathcal{K}}_w = \hat{p}/2$. We obtain good agreement between the simulation results and the theoretical predictions for both $\hat{\kappa}_w$ and $\hat{M}_w$ (Fig.~\ref{fig:mechanics_press_tubes}b). 
In a previous work, we developed a model to predict the onset of wrinkling by considering the cross-section deformation and the small but non-zero values of $N_c$~\citep{bending_luyis}. This allows for a precise determination of $\kappa_w$ and $M_w$ prior to the onset of wrinkling, in quantitative agreement with the results of the simulations at the lowest pressures. However, in contrast with TFT, the predictive power for the mechanical response beyond the wrinkling instability is weak.  A detailed description of the model and its comparison with TFT and with simulation results can be found in Sec. VII of the Supplemental Material \citep{sm_ref}.
\par
Following the characterization of the mechanical properties, we seek to determine the geometrical properties of the wrinkles. The amplitude of the wrinkles, $A$, is very small compared with $R$ (Fig.~\ref{fig:wrinkles_general}). Fourier analysis reveals that at low $\kappa/\kappa_w$ the pattern of wrinkles shows a unique wavelength. This is observed for both high and low $p$ \citep{sm_ref}. As $\kappa/\kappa_w$ increases, we observe the sudden appearance of a multi-wavelength state (MWS), where evenly spaced high amplitude wrinkles are surrounded by minor wrinkles (Fig.~\ref{fig:wrinkles_general}). Although this occurs for both low and high $p$, the spatial distribution of the wrinkles varies with pressure. For low $p$, the wrinkles with the largest amplitude are flanked by wrinkles of smaller amplitude. The amplitude of the small wrinkles quickly decays with the distance to the major wrinkle \citep{sm_ref}. This configuration resembles the profile of wrinkles observed in compressed thin sheets \citep{cerda_science_wrinkling, diamant_wrinkles_patterns}. As $p$ increases, the space between the major wrinkles shows complex patterns of high frequency and small amplitude wrinkles (Fig. \ref{fig:wrinkles_general}b and Supplemental Movies S1-S3 \citep{sm_ref}). 
\par
\begin{figure}[t!]
    \centering
    \includegraphics[width=0.49\textwidth]{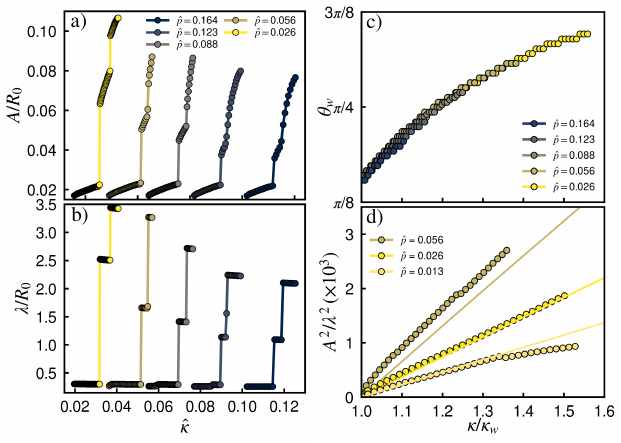}
    \caption{Geometrical properties of the wrinkles. a) Wrinkle amplitude $A$ as a function of curvature $\hat{\kappa}$ after the wrinkling transition.  b) Characteristic wavelength $\lambda$ of the largest amplitude wrinkles for different pressures. The sudden jumps correspond to the transition to the MWS. c)  Extent of the wrinkles in the sinusoidal regime measured along the polar angle $\theta_w \equiv \pi - \theta_0$ as a function of curvature. d) $A^2/\lambda^2$ relation as a function of curvature in the sinusoidal regime. The continuous lines are the predictions from Eq.~\ref{eq:amp_lambda_sq}.}
    \label{fig:wrinkle_properties}
\end{figure}

 
In light of the relatively complex patterns observed for large $\kappa$, we focused our analysis on the major wrinkles with the largest amplitude. Sustained bending after the wrinkling instability results in a slow and sublinear increase of $A$ with $\kappa$ (Fig.~\ref{fig:wrinkle_properties}a). In the sinusoidal regime we do not observe a dependency of $A$ on pressure and for every $p$, the transition to the MWS occurs when $A(\kappa)/R_0 \approx 0.02$. After this transition, the amplitude of the wrinkles increases suddenly, and additional bending results in a faster increase of $A$ compared with that in the purely sinusoidal state. 
\par 
We observe that the wavelength in the sinusoidal regime $\lambda_\mathrm{sin}$ does not change with bending (Fig.~\ref{fig:wrinkle_properties}b). Despite the similarities in wrinkle morphology, the trend observed differs from that of confined thin sheets, where there is a consistent decrease of $\lambda$ when compressed, though consistent with the theoretical predictions under a linear elasticity framework \citep{brau_mls}. Furthermore, we observe that $\lambda_\mathrm{sin}$ decreases with pressure, while the compressive stresses at the onset of wrinkling are smaller for higher pressures (Fig.~\ref{fig:mechanics_press_tubes}c) \citep{sm_ref}. Interestingly, the MWS state contains two substates. The transitions are marked by an abrupt increase of $\lambda$, coincident with a rapid increase in $A$. Once in a given state, $\lambda$ remains mostly constant with bending. Following the transition from the sinusoidal regime we observe a substate with a small wavelength $\lambda_1 > \lambda_\mathrm{sin}$ (Fig.~\ref{fig:wrinkles_general} and Fig.~\ref{fig:wrinkle_properties}b). Sustained bending results in an abrupt transition to a substate with larger wavelength ($\lambda_2 > \lambda_1 > \lambda_\mathrm{sin}$). At the largest pressures, this transition is marked by an increase of the amplitude $A$ for some of the wrinkles simultaneous with a decrease of $A$ for other wrinkles until their eventual disappearance (Supplemental Movie S3, \citep{sm_ref}).
\par
From examination of the wrinkles in the sinusoidal regime, we notice that the wrinkles do not only grow in amplitude but also in their extent along the circumferential direction (Supplemental Movies S1-S3, \citep{sm_ref}). We determine the polar angle $\theta_w$ that results in the maximum curvature along the circumferential direction as a proxy for the extent of the wrinkles, where $\theta_w \equiv \pi - \theta_0$ \citep{sm_ref}. We find that $\theta_w$ increases sublinearly with $\kappa$, resembling the response found for the wrinkle amplitude (Fig.~\ref{fig:wrinkle_properties}d). We observe the same trend for all pressures and, remarkably, $\theta_w>0$ at $\kappa_w$. The qualitative response of $\theta_w$ with $\kappa$ is in accord with that predicted by TFT (Supplemental Material Sec.~IV \citep{sm_ref}). 
We observe that larger pressures transit to the MWS at lower $\theta_w$, i.e., the transition to the multi-wavelength mode does not rely on a characteristic pressure-independent wrinkle width. 
\par
Intuitively, wrinkling is a mechanism preventing compressive stresses. Using an axial inextensibility condition for the inner-most area, building on the work of Ref.~\citep{mahadevan_inext_wrinkle_analysis} we determine the following ratio between $A$ and $\lambda$ (Supplemental Material Sec. VIII, \citep{sm_ref}):
\begin{equation}
    \label{eq:amp_lambda_sq}
    \frac{A^2}{\lambda^2}= \frac{2 \sin(\theta_w)}{\pi^2 \theta_w}(\hat{\mathcal{K}}-\hat{\mathcal{K}}_w)\;, 
\end{equation}
In contrast to \citep{mahadevan_inext_wrinkle_analysis}, we limited the inextensibility condition to the wrinkled area defined by $\theta_w$. The expression captures well the relation between the wrinkles amplitude (Fig.~\ref{fig:wrinkle_properties}a) and wavelength (Fig.~\ref{fig:wrinkle_properties}b) in the sinusoidal regime as a function of the bending curvature (Fig.~\ref{fig:wrinkle_properties}d).
\par
\emph{Conclusions}.---We have developed a computational model to precisely characterize the mechanics and geometry of pressurized tubes subjected to bending. Using a simplified theoretical framework relying on tension-field theory, we propose a model to characterize the torque-curvature curves beyond the wrinkling transition. We discover that the wrinkled state evolves with bending, encompassing first a purely sinusoidal regime which is subsequently followed by a multi-wavelength state anticipating the final buckling of the tube in a kinked configuration. 
In the sinusoidal regime, wrinkles undergo not only an increase in amplitude but also a concurrent expansion of their extent along the circumferential direction, $\theta_w$. The value of $\theta_w$ decreases with pressure at the transition to the multi-wavelength state, while $A$ is constant for all $p$. This suggests that a characteristic wrinkle amplitude is the main geometrical property triggering the transition to the multi-wavelength mode and the subsequent kinking of the tube. 
\begin{acknowledgments}
We acknowledge Benny Davidovitch for insightful discussions and L. Mahadevan for pointing out Ref.~\citep{mahadevan_inext_wrinkle_analysis}. This work was supported by Volkswagen Stiftung (A.A. and U.G.) and the Harvard Quantitative Biology Initiative and Grant NSF-1806818 (L.Q.). A.A. thanks the Clore Center for Biological Physics for their support. 
\end{acknowledgments}

\bibliographystyle{apsrev4-1} 
\bibliography{references}

\end{document}




\title{Supplementary Information\\Mechanics and wrinkling patterns of pressurized bent tubes}
\author{César L. Pastrana}
\affiliation{Physics of Complex Biosystems, Technical University of Munich, 85748 Garching, Germany}
\author{Luyi Qiu}
\affiliation{John A. Paulson School of Engineering and Applied Sciences, Harvard University, Cambridge, Massachusetts 02138, USA}
\author{John W. Hutchinson}
\affiliation{John A. Paulson School of Engineering and Applied Sciences, Harvard University, Cambridge, Massachusetts 02138, USA}
\author{Ariel Amir}
\affiliation{John A. Paulson School of Engineering and Applied Sciences, Harvard University, Cambridge, Massachusetts 02138, USA}
\affiliation{Department of Physics of Complex Systems, Weizmann Institute of Science, 7610001 Rehovot, Israel}
\author{Ulrich Gerland}
\affiliation{Physics of Complex Biosystems, Technical University of Munich, 85748 Garching, Germany}


\maketitle

\tableofcontents

\makeatletter
\let\toc@pre\relax
\let\toc@post\relax
\makeatother

\newpage
\listoffigures

\section*{Notation}
\begin{tabular}{ l@{\hskip 2em}l } 
\textbf{Symbol} & \textbf{Parameter}\\
 $p$ & Pressure \\
 $E$ & 3D Young's modulus\\
 $\nu$ & Poisson's ratio\\
 $t$ & Shell thickness\\
 $R_0$ & Undeformed radius of the tube\\ 
 $R$ & Radius of the tube (pressurized)\\ 
 $L_0$ & Undeformed length of the tube\\
 $L$ & Length of the tube\\
 $M$ & Torque\\
 $N$ & Stress resultants\\
 $\varepsilon$ & Strains\\
 $\mathcal{K}$ & Curvature of the medial axis\\
 $\kappa$ & Curvature of the innermost (concave) side of tube\\
 $A$ & Amplitude of the wrinkles\\
 $\lambda$ & Wavelength of the wrinkles\\
 $\theta_0$ & Region span by the wrinkles (polar coordinate)
\end{tabular}
\section*{Supplementary Movies}
\begin{tabularx}{\linewidth}{@{}>{\bfseries}l@{\hspace{.5em}}X@{}}
\textbf{Supplementary Movie 1 } & Bending of a pressurized tube at $\hat{p} = 0.026$. The surface is colored according to Gaussian curvature, $\kappa_G$ \citep{meyers_diff_geo_mesh}. The onset of wrinkling occurs at $t =$ 3 s. The first substate of the multi-wavelength wrinkling mode occurs at $t = $ 10 s and the transition to the second substate with a larger wavelength occurs at $t = $  20 s. Note that the curvature of the tube is not a linear function of time to highlight the different states.\\
\textbf{Supplementary Movie 2 } & Bending of a pressurized tube at $\hat{p} = 0.088$.  The onset of wrinkling occurs at $t =$ 5 s. The first substate of the multi-wavelength wrinkling mode occurs at $t = $ 13 s and the transition to the second substate with a larger wavelength occurs at $t = $  19 s.   \\
\textbf{Supplementary Movie 3 } & Bending of a pressurized tube at $\hat{p} = 0.164$.
The onset of wrinkling occurs at $t =$ 6 s. The first substate of the multi-wavelength wrinkling mode occurs at $t = $ 11 s and the transition to the second substate with a larger wavelength occurs at $t = $  15 s.
\end{tabularx}
\newpage

\section[Details of the computational model]{Details of the computational model}
\subsection{Initial configuration}
The main body of the tube is constructed by positioning particles on the surface of a cylinder describing a perfectly-ordered equilateral triangular mesh. The tube includes two lids to close the volume. The lids are constructed by including veritices with progressively reduced radius and organized to have an hexatic arrangement describing (almost) equilateral triangles. The geometrical details of the initial configuration are shown in Table \ref{tab:table_parameters}. The mesh data is obtained from the vertices' coordinates by calling to the Advancing Front Surface Reconstruction routine on the Computational Geometry Algorithms Library (CGAL).
\par
We used a  circumferentially aligned mesh because of its easiness and accuracy for the determination of both the mechanical and geometrical parameters. Similar results for the mechanical properties and the features of the wrinkles are obtained by considering a mesh with a long axis alignment (Figure \ref{fig:long_axis_mesh}).

\subsection{Model description and energy minimization}
\section{Model}
\begin{figure}[b!]
    \begin{center}
        \includegraphics[width=0.9\textwidth]{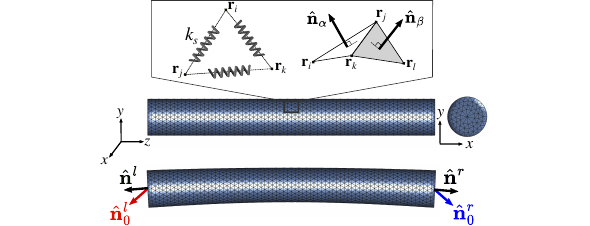}\\
    \end{center}
    \justifying
     Computational model. The cylindrical shell is modeled as triangulated mesh with stretching and bending potentials. Torque is induced on the tube by considering an energetic penalty for the miss-alignment of the normal vectors of the lids with respect to a target vector.
    \label{fig:intro_model}
\end{figure}
We study a tube of unpressurized radius  $R_0$ and length $L_0$, described by a triangulated mesh with two lids that close the volume of the tube. The total energy of the system is given by $\mathcal{E} = \mathcal{E}_\textrm{int} + \mathcal{E}_\textrm{ext}$, where $\mathcal{E}_\textrm{int}$ is the internal (mechanical) energy of the tube and $\mathcal{E}_\textrm{ext}$ is an external potential employed to bent the tube.\par 
The internal energy of the pressurized shell is given by:
\begin{equation}
    \mathcal{E}_\textrm{int} = \mathcal{E}_\textrm{b} + \mathcal{E}_\textrm{b} - pV, \label{eq:energy_int}
\end{equation}
where $\mathcal{E}_\textrm{s}$ is a stretching energy, $\mathcal{E}_\textrm{b}$ is a bending energy penalizing out-of-plane deformations of the shell, $p$ is the pressure difference and $V$ the volume enclosed by the shell \citep{pressure_induced_paper}. 
We  calculate the volume $V$ enclosed by a non-convex polyhedron using the methods described in \citep{kajiya_non_convex_polyhedron}.
\par
The stretching energy of the mesh h is described by the harmonic potential:
\begin{equation}
    \mathcal{E}_\textrm{s} = \frac{k_s}{2}\sum_{\langle i,j\rangle} (r_{ij} - r_{ij}^0)^2,\label{eq:energy_stretching}
\end{equation}
The sum runs over all pairs of connected vertices $i,j$ and $r_{ij} =\norm{\mathbf{r}_i - \mathbf{r}_j}$, where $\mathbf{r}$ are the coordinates of the vertices. The surface bending potential is given by
\begin{equation}
    \mathcal{E}_\textrm{b} = k_b\sum_{\langle \alpha,\beta\rangle}(1 - \mathbf{\hat{n}}_\alpha\cdot\mathbf{\hat{n}}_\beta),
\end{equation}
where the sum is over pairs of triangles $\alpha,\beta$ sharing an edge, and $\mathbf{\hat{n}}$  are their respective normal vectors. 
\par
This model maps to an (equivalent) continuous elastic material with 2D Young modulus  $Y = \frac{2}{\sqrt{3}}k_s$ and Poisson ratio $\nu = 1/3$ \citep{seoung_nelson_flexible_membranes}. The 2D Young modulus $Y$ is related to the 3D Young modulus $E$ by $Y=Et$, with $t$ the thickness of the shell. The discrete bending stiffness $k_b$ is assigned considering its relation with the continuum flexural rigidity $B$, given by $B= \frac{\sqrt{3}}{2}k_b$ \citep{gompper_kroll_chaper_membranes, seoung_nelson_flexible_membranes}. The value for $B$ is related to $E$ and $t$ via \citep{landau_lifshiftz_elasticity}:
\begin{equation}
    B = \frac{Et^3}{12(1-\nu^2)}.
\end{equation}
The mechanical parameters employed are shown in Table \ref{tab:table_parameters}. We noticed that the response of the mesh is only linear for low pressures (low strains). Then, we correct $E$ and $\nu$ to account for the non-linear response of the mesh. See Section \ref{sec:estim_params}\ref{ssec:corr_non_lin_mesh} for the details of the procedure.
\par 
The external potential is $\mathcal{E}_\mathrm{ext} = \mathcal{E}_\tau + \mathcal{E}_\mathrm{CM}$. The component $\mathcal{E}_\tau$ is included to induce the bending of the tube and is given by:
\begin{equation}
\mathcal{E}_\tau = k_{\tau}\sum_{ i\in\substack{\text{left}\\\text{lid}} } (1 - \mathbf{\hat{n}}_i\cdot\mathbf{\hat{n}}^l_0) + k_{\tau}\sum_{ i\in\substack{\text{right}\\\text{lid}} } (1 - \mathbf{\hat{n}}_i\cdot\mathbf{\hat{n}}_0^r) 
\label{eq:energy_ext}
\end{equation}
where the summation runs over all the triangles of the right (or left) lid and $\mathbf{\hat{n}}_i$ is the normal vector of the triangle $i$. The term $\mathbf{\hat{n}}^{l,r}_0$ is the preferred orientation vector for the right or left lid, $r$ and $l$ respectively.
\par
The term $\mathcal{E}_\mathrm{CM}$ is used to enforce the alignment of the two lids in the plane normal to the long axis of the tube 
\begin{equation}
    \mathcal{E}_\mathrm{CM} = \frac{k_\mathrm{CM}}{2}\norm{\mathbf{R}_{\mathrm{CM},x}^l  - \mathbf{R}_{\mathrm{CM},x}^r}^2
     + \frac{k_\mathrm{CM}}{2}\norm{\mathbf{R}_{\mathrm{CM},z}^l  - \mathbf{R}_{\mathrm{CM},z}^r}^2,
\end{equation}
where $\mathbf{R}_{\mathrm{CM},x}^{l,r}$ is the center of mass of the left/right lid and $x,z$ indicates the Cartesian coordinate. We found that this potential has no influence in the mechanical and geometrical response. However, in its absence a single kink is formed anywhere along the distance of the tube.
\par
The approach described resembles the action of \emph{virtual hands} to induce bending. The force exerted on the lids is adjusted with the constant $k_{\tau}$, given by $k_\tau := K_\tau k_{s}l_0^2$, where $l_0$ is the average bond length of the mesh. $K_\tau$ is a free parameter determined empirically to achieve the target curvatures of the tube. Similarly, $k_\mathrm{CM} := k_sK_\mathrm{CM}$. Notably, this procedure to bend the tube does not impose any constraint along the axial direction. The routine to bend the tube is as follows. The simulations start by first pressurizing the tube to the target pressure $p$ by progressively increasing the pressure in small steps from $p=0$. Next, to induce the bending of the tube, the preferred orientation vectors of the lids $\mathbf{\hat{n}}^{l,r}_0$ are tilted in a step-wise fashion. For instance, $\mathbf{\hat{n}}^r_0 =  0\mathbf{\hat{i}} - \sin(s\delta\theta)\mathbf{\hat{j}} + \cos(s\delta\theta)\mathbf{\hat{k}}$, where $s$ is the step number and $\delta\theta$ is a defined change in the angular position.  Each step, pressurization and bending, is followed by energy minimization. 
\par 
We use a non-linear conjugate gradient algorithm to find the configuration of the system with minimum total energy $\mathcal{E}$ using the the secant-method for line search \citep{nocedal_and_wrigth}. The tolerance criterion to escape the minimization loop relies on the average per-vertex force and the precise value is adjusted empirically for the different pressures. We consider explicit analytical expressions for all the gradients. The source code is fully written in C++. We use OpenMP to parallelize the calculations of the per-vertex gradients as well as vector operations involved in the minimization routine. This significantly improves the execution run-time for the large number of vertices employed, a requirement to appropriately characterize the wrinkles.  Importantly, our mesh model is \emph{idealized} because it does not include long range interactions. Since the surface can intersect with itself near the kinks, further bending of the kinked state would not be physical, and we exclude them from any analysis.
\par 
The compiled code, the sources and instructions of use are freely available on the public repository in \citep{github_gerland}.

\begin{table}[b!]
\begin{center}
    \begin{tabular}{|l l|}
        \hline\hline
         \textbf{Parameter} &  \textbf{Value} \\
        \hline 
         \textbf{Tube configuration}& \\
         \hline
         Tube radius (unpressurized), $R_0$ & 0.250 $\mu$m\\
         Tube length (unpressurized), $L_0$ & 2.0-12.0 $\mu$m\\
         Number of vertices, $N$ & 65,000 - 113,117\\
         \hline
         \textbf{Mechanical parameters} &\\  
         \hline
         Young modulus (3D), $E$ & 50 MPa\\
         Shell thickness, $t$ & 2.0 nm\\
         Pressure, $p$ & 0.025-0.6 atm \\
     \hline\hline
     \end{tabular}
     \end{center}
     \caption[Simulation parameters]{Simulation parameters in real units. The mechanical parameters were selected to be within the ranges observed for Gram negative bacteria \citep{Beveridge_afm_Youngs,shaevitz_afm_turgor}.}
     \label{tab:table_parameters}
\end{table}    
\section{Estimation of mechanical and geometrical properties} \label{sec:estim_params}
\subsection{Curvature, $\kappa$}
We find the curvature at the bottom side of the tube $\kappa$ from the medial axis curvature $\mathcal{K}$. First, we find a set of particles defining the medial axis of the tube $\mathbf{x}$ by taking the center of mass of consecutive sections along the tube. Then, the $\mathbf{x}$ coordinates are smoothed considering a moving average between consecutive points. 
\par 
A discrete local tangent vector is given by $\mathbf{t}_i = \mathbf{x}_{i+1} - \mathbf{x}_{i}$, where the nodes are separated by $s_i = \lVert\mathbf{x}_{i+1} - \mathbf{x}_{i}\rVert$. From the tangent vectors a local (unsigned) curvature of the medial axis is obtained as:
\begin{equation}
    \mathcal{K}_i = \left\lVert\frac{ \mathbf{\hat{t}}_{i+1} - \mathbf{\hat{t}}_{i}}{s_{i+1}- s_i}\right\rVert.
\end{equation}
Considering the local minimum radius of the tube $R_y$, the local curvature at the inner most side of the tube, is then
\begin{equation}
\kappa(s) = \frac{\mathcal{K}}{1 - \mathcal{K}R_y},
\label{eq:mean_kappa}
\end{equation}
where $\mathcal{K}=\mathcal{K}(s)$ and $R_y = R_y(s)$ since in absence of pressure, the curvature is not homogeneous along the contour length of the tube (see Section \ref{sec:length_dependency}\ref{ssec:non_press_length} and  Figure \ref{fig:geometry_length_non_press}). Then, we consider the average value of $\kappa$ on the 10\% central region of the tube, where the curvature is maximal.
\par 
In the pressurized configuration, where the curvature is mostly constant everywhere along the contour of the tube (Section \ref{sec:length_dependency}\ref{ssec:press_length}; Figure \ref{fig:geometry_length_non_press} and Figure \ref{fig:cross_section_radii_press}), we used an alternative method. From the the geometry of a curved tube with an angle $\gamma$ between the lids and the $z$ axis ($\cos\gamma = \langle\mathbf{\hat{n}}_i\cdot\mathbf{\hat{k}}\rangle_N$, where the average is over the $N$ particles of the lids), it can be shown that:
\begin{equation}
\kappa = \frac{2\gamma}{L - 2R\gamma},
\label{eq:kappa_from_lids}
\end{equation}
where $R$ and $L$ are the pressurized radius and lenght, respectively. This method to calculate an average $\kappa$ is reliable after the occurrence of the wrinkling transition, where the highly curved regions near the wrinkles results in a large bias in the estimation of the mean curvature using the method based on Equation \ref{eq:mean_kappa} . Therefore, we opted for Equation \ref{eq:kappa_from_lids} for the analysis of the complete response of the tube, including the wrinkled state (Figures 4 and 5 in the main text). This method, however, slightly overestimates the curvature at the highest pressures, as a result of deformation of the lids. Thus, we resort to the more precise values obtained from Equation \ref{eq:mean_kappa} for the determination of $\kappa_w$ (Figure 6a in the main text).

\subsection{Torque, $M$}
The external energy $\mathcal{E}_\textrm{ext}$ employed to bent the tube (Equation \ref{eq:energy_ext}) can be rewritten as:
\begin{equation}
\mathcal{E}_\textrm{ext} = k_{\tau}\sum_{ i\in\substack{\text{left}\\\text{lid}} } (1 - \cos\phi_l) + k_{\tau}\sum_{ i\in\substack{\text{right}\\\text{lid}} } (1 - \cos\phi_r) \label{eq:energy_ext_torque_2},
\end{equation}
where $\phi$ are the angles formed between the normal vector of the lids and the target normal vector. The torque is, per definition, $M = \frac{\partial E}{\partial\phi}$. Then, considering both lids:
\begin{equation} \label{eq:torque_ext}
\begin{split}
M &= \frac{1}{2}\left( \frac{\partial \mathcal{E}_\textrm{ext}}{\partial \phi_l} + \frac{\partial \mathcal{E}_\textrm{ext}}{\partial \phi_r} \right)\\
&= \frac{k_\tau}{2}\left(\sin\phi_l + \sin\phi_r\right).
\end{split}
\end{equation}

Alternatively, the torque can be determined from the mechanical energy of the tube:
\begin{equation}
M = \frac{\partial (\Delta \tilde{E}_{int})}{\partial \kappa}, \label{eq:torque_dEdk}
\end{equation}
where $\Delta \tilde{E}_{int} = \tilde{E}_{int}(\kappa) - \tilde{E}_{int}(0)$ is the difference of the internal energy \emph{per length} between the bent and the unbent pressurized configurations.
We calculate the internal energy of the shell $\mathcal{E}_\textrm{int} = E_s + E_b  - pV$, excluding from the terminal regions of the tube with the objective of minimizing artifacts associated to the lids. The torque is then calculated by obtaining the numerical derivative of the energy with respect to the curvature (Equation \ref{eq:torque_dEdk}), with $\kappa$ determined as described in the previous section. 
\par 
Both methods provide equivalent results (Figure  \ref{fig:torque_estimation}). Nevertheless, calculating the torque from Equation \ref{eq:torque_dEdk} is computationally expensive and prone to noise, specially after the wrinkling transition, due to the numerical derivative over $\kappa$. Consequently, all the results shown in this work used Equation \ref{eq:torque_ext} for the determination of $M$.

\subsection{Stress resultants, $N_{z}$}
The stresses shown in Figure 7 of the main text are the average of the stresses determined on the bottom-most part of the tube. The method described in this section is nonetheless valid to find per-vertex stress resultants of triangulated surfaces at any location. \par 
First, we determine the normal vector $\mathbf{\hat{n}}$ of the given vertex of interest. The normal vector is obtained from the triangles in the adjacent neighbour to the main vertex: 
\begin{equation}
 \mathbf{n}_i = \sum_{t=1}^Q (\mathbf{r}_{1,t} - \mathbf{r}_i)\times(\mathbf{r}_{2,t} - \mathbf{r}_i),
\end{equation}
where $Q$ is the coordination number, $\mathbf{r}_{i}$ are the coordinates of the main particle and $\mathbf{r}_{1,t}$ and $\mathbf{r}_{2,t}$ are the coordinates of the other two particles defining the triangle $t$. More precise estimates of the local normal vector can be obtained by weighting the normal vector of each triangle by the corresponding local Voronoi area
\begin{equation}
S_t = \frac{1}{8}(\norm{\mathbf{r}_1 - \mathbf{r}_i}^2\cot\angle\mathbf{r}_2 + \norm{\mathbf{r}_2 - \mathbf{r}_i}^2\cot\angle\mathbf{r}_1) ,
\end{equation}
where $\angle$ is the angle at the indicated vertex. For additional details see \cite{meyers_diff_geo_mesh}.
\par 
Having defined the local normal vector,  we project the coordinates of the $Q$ vertices of the first-ring neighbourhood of the particle $\mathbf{r}_i$ on the local plane $\mathbf{r}_j^{\text{proj}} = \mathbf{r}_{j,i} -(\mathbf{r}_{j,i}\cdot\mathbf{\hat{n}}_i)\mathbf{\hat{n}}_i$. From now on, and with the objective of keeping the notation brief, the reference to the coordinates $\mathbf{r}$ states for the  coordinates after the plane projection.
\par 
We express the coordinates of all the particles in terms of a local and arbitrary basis.  The first basis vector is given by an edge connecting one of the vertices with the main vertex: $\mathbf{e}_1 = \mathbf{r}_i - \mathbf{r}_0$. A second basis vector can be obtained as $\mathbf{e}_2 = \mathbf{e}_1 \times \mathbf{n}$. Combining these two basis vectors with the normal vector allows to obtain the local coordinate basis, defined as  $\mathbf{B}  = [\mathbf{\hat{e}}_1, \mathbf{\hat{e}}_2, \mathbf{\hat{n}}]$. Since the coordinate system here defined is orthogonal, we can express the coordinates of a given point
on the surface in terms of this basis as $\mathbf{r}' = \mathbf{B}^{\text{T}}\mathbf{r}$. 
\par 
Finally, the full stress tensor of particle $i$ is calculated as \citep{virial_theorem_md}:
\begin{equation}
 N_i = \frac{1}{S}\sum_{\left<i,j \right>}\mathbf{F}_{i,j}\otimes\mathbf{r}_{i,j}',
\end{equation}
where $\otimes$ indicates the dyadic product and $\mathbf{F}$ is the resulting force on the edge connecting vertices $i,j$. The total area of the patch is given by the sum over the Voronoi areas of the $n_t$ triangles around particle $i$:  $S = \sum_{t=1}^{n_t} S_t$. The range of indexes of the tensor are $\alpha = (1, 2, 3)$ and $\beta =(1, 2, 3)$. Plane projection and the change of basis results in $N^{\alpha\beta} = 0$ for every component of tensor having $\alpha = 3$ or $\beta = 3$.\par 
The stress tensor obtained is defined on the arbitrary basis given by the connectivity.  Thus, we need to change to a target basis of choice $\mathbf{B} = [\mathbf{m}_1, \mathbf{m}_2, \mathbf{m}_3 ]$ following the tensor transformation law, $\mathbf{N}' =  \mathbf{Q} \mathbf{N} \mathbf{Q}^\textrm{T}$, where the matrix $\mathbf{Q}$ is given by
\begin{equation}
 \mathbf{Q} = 
 \begin{bmatrix} 
  \mathbf{m_1}\cdot\mathbf{\hat{e}_1} & \mathbf{m_1}\cdot\mathbf{\hat{e}_2} & \mathbf{m_1}\cdot\mathbf{\hat{n}}\\
  \mathbf{m_2}\cdot\mathbf{\hat{e}_1} & \mathbf{m_2}\cdot\mathbf{\hat{e}_2} & \mathbf{m_2}\cdot\mathbf{\hat{n}}\\
  \mathbf{m_3}\cdot\mathbf{\hat{e}_1} & \mathbf{m_3}\cdot\mathbf{\hat{e}_2} & \mathbf{m_3}\cdot\mathbf{\hat{n}}\\
 \end{bmatrix}.
\end{equation}
In our particular case, the target basis are those giving the orientation along the axial and circumferential directions of the tube and $\mathbf{m}_3 \equiv \mathbf{\hat{n}}$. The basis vector in the axial direction is determined considering the tangent vectors of the vertices defining the bottom-most part of the tube: $\mathbf{m}_1 = \mathbf{r}_{i+1} - \mathbf{r}_{i}$, excluding the component along the $x$ axis, which arises as a consequence of the triangular geometry of the mesh. The basis vector in the circumferential direction can be obtained as $\mathbf{m}_2 = \mathbf{m}_{1}\times  \mathbf{n}_{i}$.
\par 
In Figure 4 we show the stress tensor in the axial direction as the average over vertices on the bottom-most part of the tube and excluding regions near the terminal sides. Note that the tensor obtained is of mixed components $N_z^z$. In other words, we use lower indexes $N_{zz}$ for convenience of notation, without implying a covariant form.

\subsection{Extent of wrinkles, $\theta_w$}
We use the curvature along the circumferential direction $\kappa_{\theta\theta}$ as a proxy for the extent of the wrinkles. The curvature $\kappa_{\theta\theta}$ is calculated from Gaussian and mean curvatures, as described in \cite{meyers_diff_geo_mesh}.
\par
The bending of the tube results in a deformation of the cross-section to an elliptic shape. Though the eccentricity is very small (Figure \ref{fig:cross_section_radii_press}), as a consequence of the elliptic cross-section of tube, the curvature along the circumferential direction has two maxima: a peak at $\sim \pi/2$ and other caused by the wrinkled region prior to decay to non-wrinkled regions (Figure \ref{fig:extent_wrinkles}a and \ref{fig:extent_wrinkles}b). However, particularly at the largest pressures where the eccentricity of the tube is larger, the position of the first peak associated to wrinkled area is hard to resolve. To make the identification of the extent of the wrinkles easier, we subtract the response immediately before the onset of wrinkling. In mathematical terms, $\kappa_{\theta\theta}^{off}(\theta, \kappa) = \kappa_{\theta\theta}(\theta, \kappa) - \kappa_{\theta\theta}(\theta, \kappa_w)$. The angular extent of the wrinkles is given by: $\theta_w(\kappa) = \argmax_\theta \{\kappa_{\theta\theta}^{off}(\theta, \kappa)\}$. Examples for $\kappa_{\theta\theta}^{off}$ at different post-wrinkling curvatures are shown in Figure \ref{fig:extent_wrinkles}a-c. 

\subsection{Corrections for the non-linearity of the mesh} \label{ssec:corr_non_lin_mesh}
The mapping between the mesh model and linear elasticity only applies for low strains. At the pressures employed in this work the strains are moderately small ($\approx$ 0.17 at the highest pressure applied), yet significant to manifest a deviation from linearity (Figure \ref{fig:non_linearity_mesh}). We correct the Young's modulus $E$ and the Poisson's ratio $\nu$ to account for the aforementioned non-linearity of the mesh model.\par 
The starting point are the expected strains in a pressurized hollow tube:
\begin{equation}\label{eq:strin_theta_tube}
\varepsilon_{\theta\theta} = \frac{pR}{Et}\left( 1 -  \frac{\nu}{2}\right)
\end{equation}
\begin{equation}\label{eq:strin_z_tube}
\varepsilon_{zz} = \frac{pR}{Et}\left( \frac{1}{2}- \nu\right),
\end{equation}
where $\varepsilon_{\theta\theta}$ and $\varepsilon_{zz}$ are the strains in the circumferential and axial direction, respectively. These equations arise after considering that the stresses in a pressurized cylinder are $2N_{zz} = N_{\theta\theta} = pR$. Taking the derivatives with respect to pressure of Equations \ref{eq:strin_theta_tube} and \ref{eq:strin_z_tube}, we can solve for an effective Young's modulus $E_\textrm{eff}$ and Poisson's ratio $\nu_\textrm{eff}$,
\begin{equation}
E_\textrm{eff} = \frac{3R_0 \partial_p\varepsilon_{\theta\theta}}{2t(2\partial_p\varepsilon_{\theta\theta} - \partial_p\varepsilon_{zz} )}, \label{eq:E_eff}
\end{equation}
\begin{equation}
\nu_\textrm{eff} = \frac{2\partial_p\varepsilon_{zz} - \partial_p\varepsilon_{\theta\theta}}{\partial_p\varepsilon_{zz} - 2\partial_p\varepsilon_{\theta\theta}}.
\label{eq:nu_eff}
\end{equation}
The average strains on the tube are given by $\varepsilon_{\theta\theta} = (R-R_0)/R_0$ and $\varepsilon_{zz} = (L-L_0)/L_0$. To determine the $E_\textrm{eff}$ and $\nu_\textrm{eff}$ from the simulation, we find average radius $R$ and length $L$ in a region away from the terminal sides to prevent artifacts associated to the lids (Figure \ref{fig:non_linearity_mesh}). The values $\partial_p \varepsilon_{\theta\theta}$ and $\partial_p \varepsilon_{zz}$, are obtained from the derivatives of a 5th order polynomial fit to the strains at different pressures in absence of torque. The resulting $\partial_p \varepsilon_{\theta\theta} $ and $\partial_p \varepsilon_{zz}$ for each pressure is inserted in Equation \ref{eq:E_eff} and \ref{eq:nu_eff} to determine the effective Young's modulus and Poisson's ratio. 

\section{Mechanics and Geometry of unpressurized tubes}
The buckling of a tube in the absence of pressure has been widely studied during the last century \citep{calladines_brazier_chapter, brazier_main}. Therefore, we use the unpressurized condition as a proof-of-principle of the ability of the computational model to produce well-established results. First, we introduce the non-dimensional curvature and torque in the so-called \emph{shell} normalization, $\bar{\kappa} :=  \frac{\kappa R}{\alpha}$ and torque $\bar{M} := \frac{M}{\alpha EtR^2}$, where $\alpha := \frac{t}{R}\sqrt{\frac{1}{12(1-\nu^2)}}$. For thin shells $\alpha \ll 1$ and in our case, and unless otherwise stated, $\alpha =$ 0.0024 in the unpressurized configuration. In light of the indicated normalization, the relation between torque and curvature is, up to second order approximation \citep{bending_luyis, karamanos_bending}:
\begin{equation}
    \bar{M} = \pi\left(\bar{\kappa} - \frac{\bar{\kappa}^3}{8}\right).
    \label{eq:torque_shell}
\end{equation}

We find that, for sufficiently long tubes ($L \gg R$), the torque-curvature response is very well described by the numerical model (Figure~\ref{fig:unpress_data}a). An extensive analysis of length dependencies is shown in Sec. \ref{sec:length_dependency}A. We find quantitative agreement between the model and the analytical theory for the full $\bar{\kappa}$ range. Moreover, our simulations show that the curvature leading to the buckling instability is in accord with the theoretical estimation, which predicts $\bar{\kappa}_w = 1.320$ for the wrinkling instability \citep{bending_luyis}. We observe that the simulation shows a slightly earlier instability, occurring at a $\sim$3\% lower $\bar{\kappa}$ than the predicted $\bar{\kappa}_w$. Adjusting the simulation parameters to produce smaller changes in $\kappa$ per iteration (e.g. reducing $k_\tau$) results in very similar $\kappa_w$ values and in every case occurring slightly earlier than the theoretical prediction.
\par 
Furthermore, we analyze the geometry of the tube during bending. It is predicted that the cross-section of the tube acquires an elliptical shape \citep{calladines_brazier_chapter, bending_luyis}. Simulations confirm the formation of an elliptical cross-section during bending and the eccentricity at the instability is in very good agreement with the theoretical predictions for the semi-major and semi-minor radii (Figure~\ref{fig:unpress_data}b, \citep{bending_luyis}).

\section{Finite length dependency}\label{sec:length_dependency}
The analytical approximations to understand the buckling phenomena rely on the assumption of an infinitely long tube with isotropic curvature and radius. In the real world, as well as in the simulations, an infinite length is not feasible. In this section we study the finite-size implications on the mechanical and geometrical properties of tubes subjected to bending by using tubes of different lengths. 

\subsection{Non-pressurized tubes}\label{ssec:non_press_length}
First, we investigate the mechanical response for different lengths in an unpressurized situation. From the torque-curvature curves we find that shorter tubes show a linear torque-curvature dependency (Figure \ref{fig:mechanics_length_non_press}a), resembling the mechanics of a solid tube \citep{landau_lifshiftz_elasticity}. In addition, the curvature at buckling (and thus the torque) is higher in short tubes than in the longer ones. As the tube length increases, the $\kappa_w$ and $M_w$ required for buckling progressively decrease, ultimately converging to the values expected from analytical theory (Figure \ref{fig:mechanics_length_non_press}b and Figure \ref{fig:mechanics_length_non_press}c).
\par 
Next, we observe that shorter tubes do not undergo elliptical deformations of the cross-section during the bending process (Figure \ref{fig:geometry_length_non_press}a). Similarly as we observe with the mechanical properties, the semimajor and semiminor radii (measured at a central region and accounting for a length of a 10\% of the total length) steadily converge to the expected radii predicted by analytical theory as the length of the tube increases (Figure \ref{fig:geometry_length_non_press}b).
Notably, we find that the radius of the tube is not constant along the contour length. We observe that, for long tubes, the resulting eccentricity is maximum  at the central region of the tube, while the terminal sides maintain a circular cross-section (Figure \ref{fig:geometry_length_non_press}c). The curvature of the tube displays a similar profile: short tubes exhibit a contour-independent curvature and longer tubes are maximally curved on the middle region and minimal at the lids (Figure \ref{fig:geometry_length_non_press}d). These results are consistent with the finite-length effects described when $\Omega := \sqrt{L^2h/R_0^3} \ll 1$, where $h:=\frac{t}{\sqrt{1-\nu^2}}$ \citep{calladines_brazier_chapter}.
\par 
In light of these findings, we decided to use tubes with $L/R_0=48$ (corresponding to $\Omega = 4.42$) for the analysis of the mechanical and geometrical response to bending in the unpressurized configuration (Figure  \ref{fig:unpress_data}). 

\subsection{Pressurized tubes}\label{ssec:press_length}
In contrast to the unpressurized case, the mechanical response of the pressurized tube is length-independent for the range of pressures studied. We do not find significant differences or length dependencies in the wrinkling and torque curvatures, even at the lowest pressures (Figure \ref{fig:mechanics_length_press}). We find a negligible effect of the length on the geometrical properties of the tube: even at the lowest pressure, the dependency of the radii and the curvature on the position along the tube is suppressed by the internal pressure (Figure \ref{fig:geometry_length_press}).
Therefore, we resort to tubes with $L/R_0 = 24$ for the analysis of the wrinkles shown in the main text. This length allowed us to appropriately measure the amplitude and wavelength of the wrinkles, keeping the number of vertices within a computationally tractable limit, $N = 113,117$ vertices. For the determination of the $A^2/\lambda^2$ relation (Fig. 4d of the main text) we increased the refinement to $N =230,534$ vertices. This was a requirement to precisely quantify $A$ near the onset of wrinkling, where the amplitude is minimal.
\par
Remarkably, we observe that for the pressures employed in this work, the cross-section remains mostly circular during bending (Figure \ref{fig:cross_section_radii_press}a). Pressure prevents the deformation of the cross-section of the tube and then higher pressures are less eccentric for a given $\kappa$. However, since the curvature necessary to buckle the tube grows with $p$, the eccentricity at the wrinkling curvature also grows with pressure (Figure \ref{fig:cross_section_radii_press}b). We find that even at very small pressures (radial strains post pressurization below 1\%), the radii of tube and the curvature are independent on the position along the contour length (Figure \ref{fig:cross_section_radii_press}b and Figure \ref{fig:cross_section_radii_press}c).


\section{Bending of pressurized cylinder: a tension field theory model}

Here we consider the pure bending of a long-capped pressurized cylindrical membrane accounting for wrinkling where axial compression occurs. 

\begin{figure}[h!]
    \centering
    \includegraphics[width=0.8\textwidth]{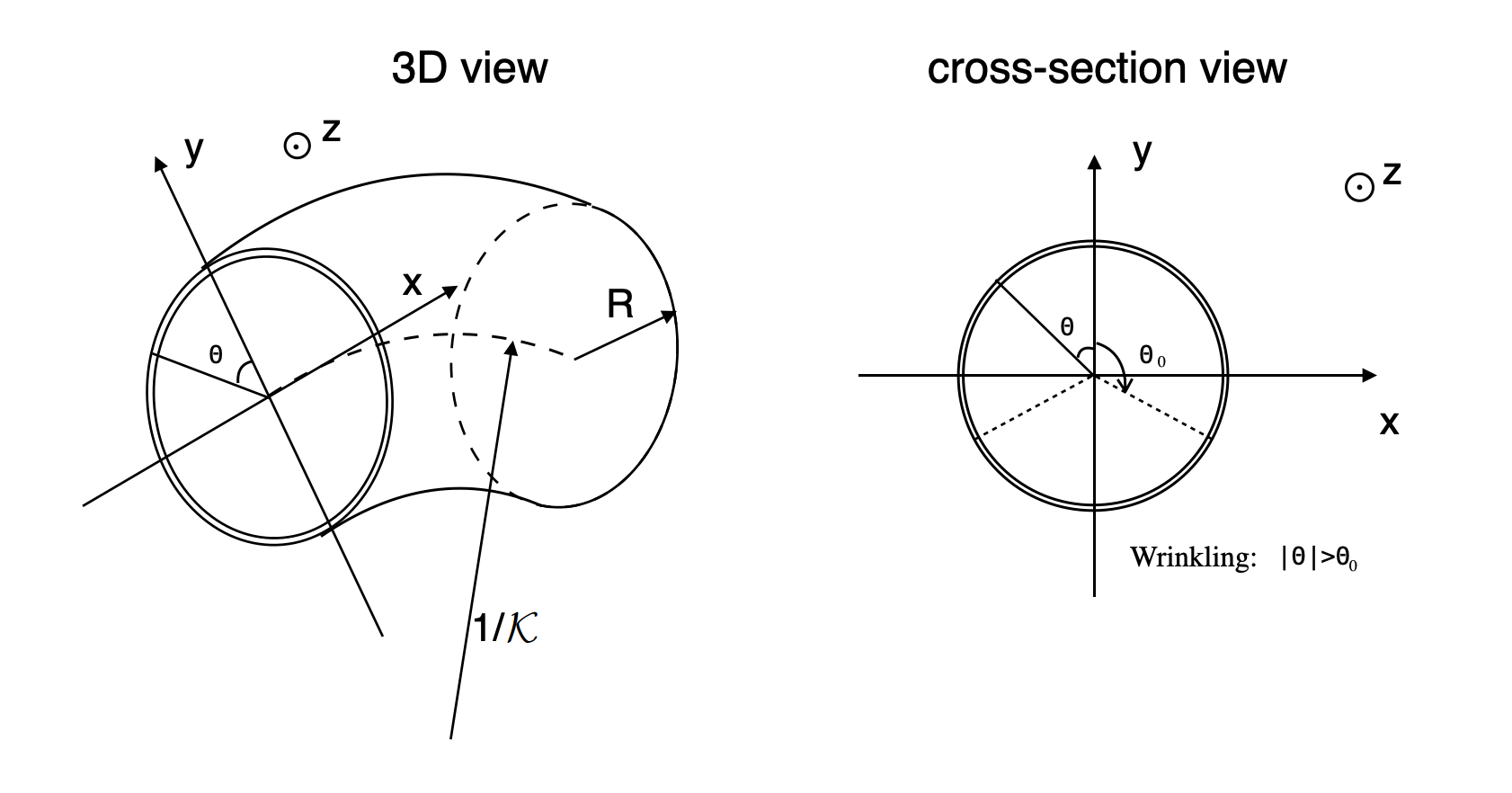}\\
    Geometry of bent tubes: notation. The long axis is defined along the length of the tube.
    \label{fig:kw_mw}
\end{figure}

The long-capped cylindrical membrane is first pressurized to $p$ which is then held
constant and the cylinder is subject to pure bending about the $x$-axis. In the modeling analysis conducted here, the strains are assumed to be small and the membrane is modeled by plane stress linear elasticity. In the region of the cylindrical membrane tube where the axial stress would be negative for the wrinkling to occur, the axial stress is taken to be zero as an approximation for the convenience of calculation. This is the source of the non-linearity and the buckling-like behavior. Prior to bending, the resultant stress in the cylindrical membrane tube subject to pressure $p$ are
\begin{equation}
 N_z = \frac{pR}{2} \;, N_\theta = pR \;,
\end{equation}
where $R$ is the radius of the tube. During pure bending at constant $p$, the axial strain satisfies $\varepsilon_z = \varepsilon_0 + \mathcal{K} z$, where we will regard the axial curvature of the cylinder center-line as imposed and its axial strain $\varepsilon_0$ as an unknown. In the wrinkled region, $\varepsilon_z$ is the axial strain averaged over the wrinkles and not the axial strain in the membrane itself, however, $\varepsilon_z$ in the wrinkled region does not play a role in the analysis.

In the unwrinkled region of the tube the linear relations relating the resultant membrane forces per length to the strains are
\begin{equation}
 N_z = \frac{Et}{1-\nu^2}(\varepsilon_z +\nu \varepsilon_\theta) \;, N_\theta = \frac{Et}{1-\nu^2}(\varepsilon_\theta +\nu \varepsilon_z) \;.
\end{equation}
Here we limit our attention to isotropic linear elasticity. 
In this analysis we assume (as an approximation) the pressure is large enough such that each cross-section remains circular and, thus, circumferential equilibrium requires $N_\theta = pR$ around the entire circumference whether wrinkling occurs or not.

The energy per unit length of the tube is the sum of the strain energy and the potential energy of the pressure:
\begin{equation}
 \Phi = \int_0^{\theta_0} N_z \varepsilon_z R d \theta + \int_0^{\pi} N_\theta \varepsilon_\theta R d \theta - p\Delta V \;,
\end{equation}
with, for $|\theta| < \theta_0$,

\begin{align}
    & N_z = Et(\varepsilon_0 + \mathcal{K} R \cos\theta) +\nu pR \;, N_\theta= pR\;, \\ 
  & \varepsilon_z = \varepsilon_0 + \mathcal{K} R \cos{\theta} \;, 
 \varepsilon_\theta = (1-\nu^2)pR/Et -\nu(\varepsilon_0 + \mathcal{K} R \cos{\theta})\;.
\end{align}
And, for $|\theta| > \theta_0$,
\begin{equation}
     N_z = 0 \;, N_\theta= pR\;,  
     \mbox{and} \;
     \varepsilon_\theta = \frac{N_\theta}{Et}\;.
     \label{eq:stress_comp_zero}
\end{equation}
And the expression for $\Delta V$ is
\begin{equation}
 \Delta V = \int_0^{\pi} 2(1+\varepsilon_z)R^2 \sin{\theta}^2 d\theta + \int_0^\pi 2R^2 \varepsilon_\theta d \theta \;.
\end{equation}

Given that the condition for boundary $\theta_0$ between wrinkled and unwrinkled regions is $N_z = 0$, we have the equation for $\theta_0$:
\begin{equation}
    \varepsilon_0 + \mathcal{K} R \cos\theta_0 = -\nu \frac{pR}{Et} \;.
    \label{eq:theta0}
\end{equation}

We then show two equivalent ways to determine the strain $\varepsilon_0$ and then the state of the system: direct approach using the force-balance equation and energy minimization with regard to $\varepsilon_0$. 

\subsection{Direct approach}
The overall equilibrium in the axial direction gives:
\begin{equation}
    \int_0^{\theta_0} 2 N_z R d \theta = \pi R^2 p\;,
\end{equation}
which results in 
\begin{equation}
    2 \mathcal{K} R (\sin{\theta_0} -\theta_0 \cos{\theta_0}) = \pi \frac{pR}{Et} \;.
    \label{eq:force_balance}
\end{equation}
after the wrinkling has happened ($\theta_0 < \pi$). The expansion of the wrinkles with the increase in curvature is shown in Figure \ref{fig:tft_wrinkle_expansion}. 
\par
The moment about the x-axis (the cross-section center) gives,
\begin{equation}
    M= \int_0^{\theta_0} 2 N_z R^2 \cos{\theta} d \theta\ = EtR^2 \left[\theta_0-\frac{\sin{(2\theta_0)}}{2}  \right] \mathcal{K} R\;.
\end{equation}
Again this is the result after the onset of wrinkling. Thus, the moment-curvature relation for fixed $p$ is determined by varying $\theta_0$ and plotting the above two relations. 

Prior to the onset of wrinkling, we won't find a point where $N_z =0$ so Equation \ref{eq:theta0} won't be applicable for this case. Then the overall equilibrium in the axial direction (Equation \ref{eq:force_balance}) should become:
\begin{equation}
    \int_0^{\pi} 2 N_z R d \theta = \pi R^2 p\;,
\end{equation}
which results in 
\begin{equation}
    Et\varepsilon_0 = \frac{pR}{2} - \nu pR.
\end{equation}
And thus we have the expression for $N_z$ and $M$ before the onset of wrinkling: 
\begin{equation}
    N_z = \frac{pR}{2} + Et\mathcal{K} R \cos{\theta}
    \label{eq:stress_pre_wrinkling}
\end{equation}
\begin{equation}
    M = \pi \mathcal{K} Et R^3.
    \label{eq:torque_tft_linear}
\end{equation}
Since we assume $N_z = 0$ for $\theta_0$, we can use Equation \ref{eq:stress_pre_wrinkling} to determine the curvature at wrinkling
\begin{equation}
    \hat{\mathcal{K}}_w = \frac{\hat{p}}{2},
    \label{eq:kappa_w_tft}
\end{equation}
where $\hat{\mathcal{K}} := \mathcal{K}R$ and $\hat{p} := pR/(Et)$.
\subsection{Energy approach}

With $p$ and $\mathcal{K}$ regarded as prescribed, $\varepsilon_0$ must minimize $\phi$ :
\begin{equation}
    0=\frac{\partial \Phi}{\partial \varepsilon_0} 
    =\frac{d}{d\varepsilon_0}\left(\int_0^{\theta_0} N_z\varepsilon_z R d \theta \right)
    +\int_{0}^{\pi}\frac{d(N_\theta \varepsilon_\theta)}{d\varepsilon_0} R d\theta -p\int_0^\pi 
    \left(2|x|R\sin\theta + 2R^2 \frac{d\varepsilon_\theta}{d\varepsilon_0}\right) d\theta \;.
\end{equation}
And we can calculate each term separately, 
\begin{align}
    \frac{d}{d\varepsilon_0}\left(\int_0^{\theta_0} N_z\varepsilon_z R d \theta \right) & = 
    \int_0^{\theta_0} (2N_z- \nu N_\theta)R d \theta + 0 \;,
    \\
    \int_{0}^{\pi}\frac{d(N_\theta \varepsilon_\theta)}{d\varepsilon_0} R d\theta & = \int_{0}^{\theta_0}- \nu N_\theta R d\theta + N_\theta \varepsilon_\theta 2R \frac{d\varepsilon_\theta}{d\varepsilon_0}\Bigr|_{\theta=\theta_0} \;, \\
    -p \int_0^\pi 
    2|x|R\sin\theta  d\theta & = -p\pi R^2 \;, \\
    -p \int_0^\pi 2R^2 \frac{d\varepsilon_\theta}{d\varepsilon_0} d\theta & = -p \int_0^{\theta_0} 2R^2 (-\nu) d\theta
    - 2R^2 p \varepsilon_{\theta} \frac{d\theta_0}{d\varepsilon_0}\Bigr|_{\theta=\theta_0} \;. \\
\end{align}
Adding all terms together, we have 
\begin{equation}
 \int_{0}^{\theta_0} 2 N_z R d \theta = p\pi R^2 \;,
\end{equation}
which is the same as the equation derived from force-balance equation. 

The moment $M$ is given by,
\begin{equation}
    M = \frac{\partial \Phi}{\partial \mathcal{K}} \;.
\end{equation}
Using the equation specifying the wrinkling boundary and the expression for $\varepsilon_0$ above, we have:
\begin{equation}
    \frac{M}{EtR^2} = \left[ \theta_0 - \frac{\sin{(2\theta_0)}}{2} \right] \mathcal{K} R \;.
    \label{eq:torque_tft_wrinkled}
\end{equation}
The resulting expression for $M$ agrees exactly with the relation derived by the direct method. The pre-wrinkling results are the same too and these can be derived by setting $\theta_0 = \pi$ in the system energy $\Phi$ .

\section{Bending of pressurized cylinder: non-linear stress-strain relation}
Here, we summarize the main findings of the model presented in \citep{bending_luyis}. Compared to the TFT model described in the main text, this model took cross-section deformations into account and used the exact expression for the critical compressive stress in Equation 7 in the main text.
\par 
In particular, the theoretical expression for the shape of the cross-section, namely the semi-major and semi-minor radii, $R_x$ and $R_y$, is given by: 
\begin{equation}
    R_x = \frac{R}{2} \int_{-\pi/2}^{\pi/2} [\cos\mu + a_2\sin(2\mu) ]d\mu 
    \label{eq:rx_integral}
\end{equation}
\begin{equation}
    R_y = \frac{R}{2} \int_{0}^{\pi} [\sin\mu + a_2\sin(2\mu) ]d\mu
    \label{eq:ry_integral}
\end{equation}
The factor $a_2$ is defined in Equation 7 in \cite{bending_luyis}:
\begin{equation}
    a_2 = -\frac{\hat{\kappa}^2}{8(\alpha^2 + \hat{p}/3)}\;,
\end{equation}
where $\alpha := \frac{t}{R}\sqrt{\frac{1}{12(1-\nu^2)}}$. Taylor expansion of the trigonometric functions in Eqs.~\ref{eq:rx_integral} and \ref{eq:ry_integral} and considering up to first order allows to write:
\begin{equation}
    \label{eq:rx}
    R_x \approx R\left(1 - \frac{2}{3}a_2\right),
\end{equation}
\begin{equation}
    \label{eq:ry}
    R_y \approx R\left(1  + \frac{2}{3}a_2\right),
\end{equation}
The eccentricity is then given by:
\begin{equation} 
    \label{eq:rxry}
    \frac{R_x}{R_y} \approx \frac{3-2a_2}{3+2a_2} \;,
\end{equation}
Equations \ref{eq:rx} and \ref{eq:ry} are derived from Equations S26 and S27 in \cite{bending_luyis}, which denotes the shape of the cross-section from the coefficients $a_n$ (for $n=1, 2,...,\infty$). For the zero pressure case, it was proven that $a_2$ is the leading term. For the pressurized case, we proved that only even terms are allowed and we assume $a_2$ is the leading term while we approximate all the other $a_n$ terms to be zero. All together leads to Eqs.~\ref{eq:rx}-\ref{eq:ry}. Interestingly, Eq.~\ref{eq:rxry} predicts a non-monotonic dependency of the cross-section shape with pressure, reaching almost no-deformation in the range of $\hat{p}$ from 10$^{-3}$ to 10$^{-1}$. Our numerical results are in good agreement with the predictions given by Eq.~\ref{eq:rxry} (Figure \ref{fig:cross_section_radii_press}b).
\par 
Accounting for the deformation of the cross-section, we obtain the following expression for the torque-curvature relation:
\begin{equation}
    \hat{M} = \pi\left[\hat{\kappa} + \frac{\hat{\kappa}^3}{\alpha^2 + \hat{p}/3} \right].
    \label{eq:torque_curv_qiu}
\end{equation}
In Figure \ref{fig:torque_kw_mw_comp}a we show the simulation results together with the predictions of \ref{eq:torque_curv_qiu} for different pressures.
\par 
We found that the dimensionless critical wrinkling curvature $\hat{\kappa}_w$ follows the equation:
\begin{equation}
8\left(\alpha^2 + \frac{\hat{p}}{3} \right)\left(\frac{\hat{p}}{2} + 2\alpha - \hat{\kappa}_w \right) + \left(\frac{2\hat{\kappa}_w}{3} -4\alpha\right)\hat{\kappa}_w^2  = 0 \,.
\label{eq:kappa_w}
\end{equation}
For pressurized thin shells such that  $\alpha \ll \hat{p}$, Equation~\ref{eq:kappa_w} can be written, to first order as, $\hat{\kappa}_w \approx \hat{p}/2$. We compare the results from computational model with the predicted $\kappa_w$ given by \ref{eq:kappa_w} and by the tension field theory model described in the previous section (Equations~\ref{eq:kappa_w} and \ref{eq:stress_comp_zero}) in Figure \ref{fig:torque_kw_mw_comp}b and Figure \ref{fig:torque_kw_mw_comp}c. We find good agreement between the simulation results and the theoretical predictions for
both $\hat{\kappa}_w$ and $M_w$. We remark the quantitative agreement at the lowest pressures between the model and the theory. 
\par 

\section{Wrinkles geometry from axial in-extensibility conditions}
Here we derive the expression for the relation between the amplitude $A$ and wavelength $\lambda$  of the wrinkles in the sinusoidal regime by considering a geometric constrain due to axial in-extensibility. We will follow an approach similar to that described by Mahadevan et al.~\citep{mahadevan_inext_wrinkle_analysis}. We assume that the radial displacement of the wrinkles is sinusoidal in the longitudinal and in the circumferential directions. Thus, in cylindrical polar coordinates $(r, \theta, z)$ (as defined in Fig. 1 in the main text), we can write down the radial displacement of the wrinkles as $u(r,\theta,z)=A\sin (2\pi z/\lambda)\cos((\pi-\theta)/\theta_w\pi/2)$.  The term $\theta_w=\pi-\theta_0$ is the extent of the wrinkles. 
\par
The axial strain of wrinkles forming on a bent tube with imposed medial axis curvature $\mathcal{K}$, compared to that of $\mathcal{K}_w$, has two contributions. One is the strain caused by the sinusoidal wrinkles and it bears the form $\frac{\sqrt{dz^2+du^2}}{dz}-1 \approx \frac{1}{2}\left(\frac{\partial u}{\partial z}\right) ^2$. 
The other comes from the imposed curvature $\mathcal{K}$ and it bears the form $R (\mathcal{K}-\mathcal{K}_w) \cos(\pi-\theta)$ under the assumption of circular cross-section with radius $R$. Then, at certain position $(r, \theta, z)$, the total axial strain $\varepsilon$ is approximately:
\begin{equation}
    \varepsilon=\frac{1}{2}\left(\frac{\partial u}{\partial z}\right) ^2 -R (\mathcal{K}-\mathcal{K}_w) \cos(\pi-\theta)\;.
\end{equation}
We express the axial in-extensibility constraint in the following form:
\begin{equation}
    0=2\int_{\pi-\theta_w} ^{\pi}\int_0^{\lambda} \left[ \frac{1}{2}\left(\frac{\partial u}{\partial z}\right) ^2 -(\hat{\mathcal{K}}-\hat{\mathcal{K}}_w) \cos(\pi-\theta) \right] dz d\theta \;,
\end{equation}
which leads to
\begin{equation}
    \frac{A^2}{\lambda^2} = \frac{2 \sin(\theta_w)}{\pi^2 \theta_w}(\hat{\mathcal{K}}-\hat{\mathcal{K}}_w)\;.
    \label{eq:amp_lambda_sq}
\end{equation}
Since $\theta_w$ is independent of pressure (Fig. 4c of the main text) the $A/\lambda$ relation is preserved for all pressures.
{\color{blue}The functional form of equation \ref{eq:amp_lambda_sq} is different to that by Mahadevan et al. ~\citep{mahadevan_inext_wrinkle_analysis}. However, we recover the same expression for $\theta_w=\pi/2$, which leads to $\frac{A^2}{\lambda^2} = \frac{4}{\pi^3}(\hat{\mathcal{K}}-\hat{\mathcal{K}}_w)$.}
\par 
To better quantify $A$ near the onset of wrinkling, where the amplitude is negligible compared with the radius of the tube, we increased the resolution the mesh to $N =230,534$ vertices to produce the data in Fig.~4d of the main text. {\color{blue}Still, we observe a progressive deviation between the numerical results and the theory as the pressure increases (Fig. \ref{fig:amp_lampbda_sq}). We attribute this deviation to the implicit loss of refinement with the expansion of the area tube $p$, reducing the accuracy on the determination of both $A$ and $\kappa$ beyond the wrinkling transition.}


\newpage
\section{Supplementary Figures}

\begin{figure}[h!]
    \centering
    \includegraphics[width=0.95\textwidth]{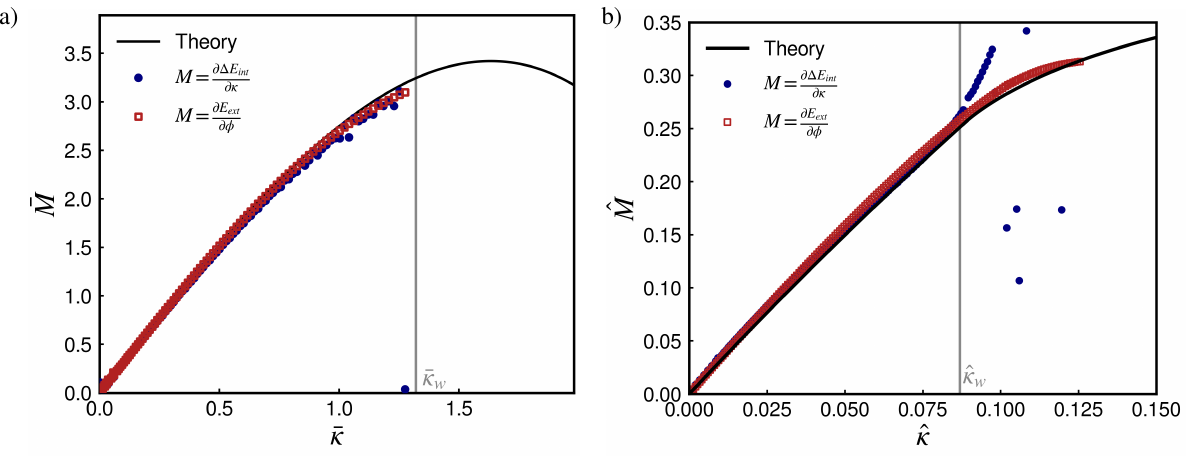}
    \caption[Estimation of torque in the computational model]{Estimation of torque in the computational model. Comparison of methodologies. a) $M-\kappa$ curves in the unpressurized configuration. b) $M-\kappa$ curves in the pressurized configuration for $\hat{p} = 0.126$.  Large noise is observed after wrinkling when the torque is calculated from the numerical derivative of the energy-curvature ($M = \partial_\kappa \Delta \tilde{E}_{int}$).}
    \label{fig:torque_estimation}
\end{figure}
\clearpage

\begin{figure}[t!]
    \centering
    \includegraphics[width=0.95\textwidth]{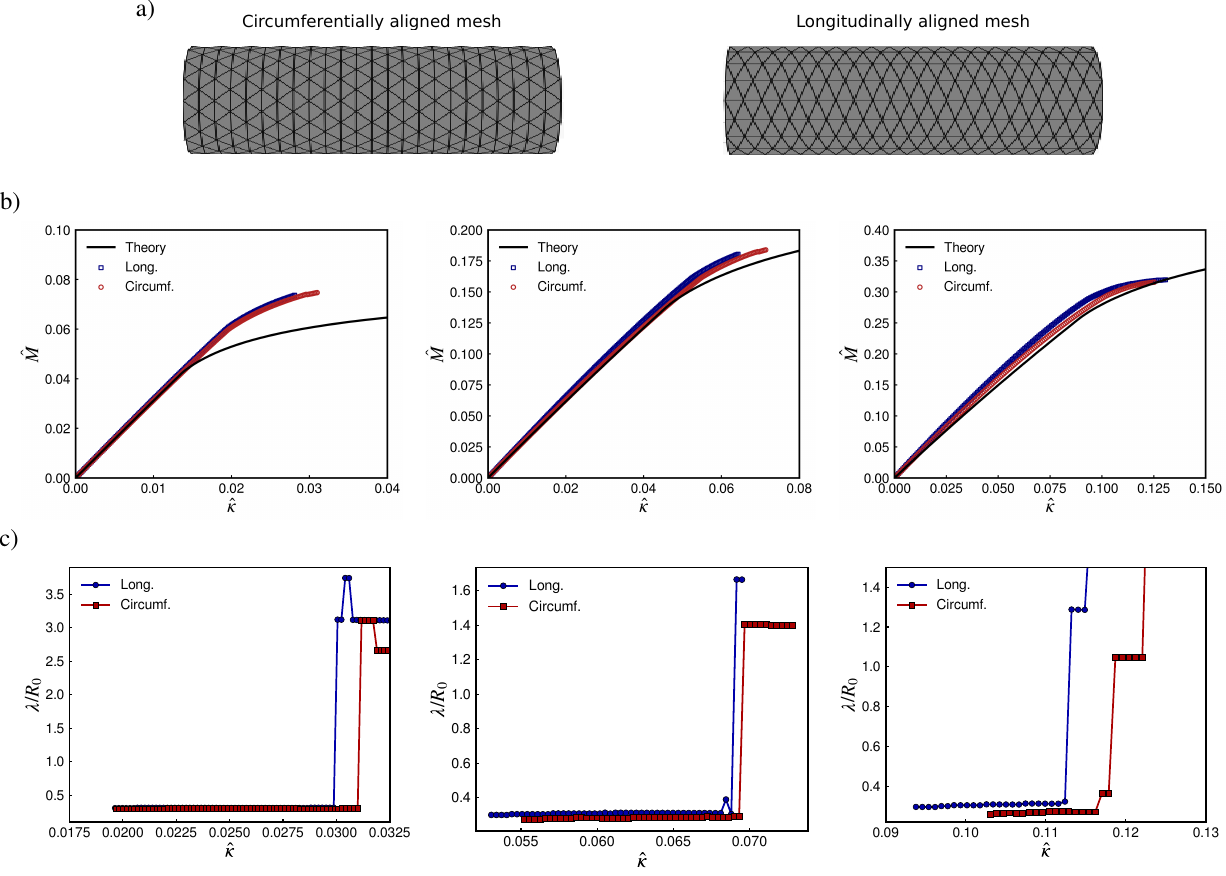}
    \caption[Longitudinally aligned meshes]{Longitudinally aligned meshes: summary of mechanical and geometrical properties a) Comparison of the different mesh geometries. b) Torque-curvature ($M-\kappa$) curves in the circumferentially aligned mesh vs. the longitudinally aligned mesh. c) Wavelength of the wrinkles $\lambda$ in circumferentially and longitudinally aligned meshes. The rescaled pressures in panels b) and c) are, from left to right, $\hat{p} = $ 0.026, 0.088 and 0.164. The correction in the effective Young's modulus $E_{eff}$ and Poisson's ratio $\nu_{eff}$ for the rescaling were obtained considering pressurization data from the circumferentially aligned configuration. The subtle differences in both $M-\kappa$ and $\lambda$ as pressure increases can consequently be attributed to not having considered specific $E_{eff}$ and $\nu_{eff}$ for the longitudinally aligned mesh.}
    \label{fig:long_axis_mesh}
\end{figure}
\clearpage

\begin{figure}[t!]
    \centering
    \includegraphics[width=0.95\textwidth]{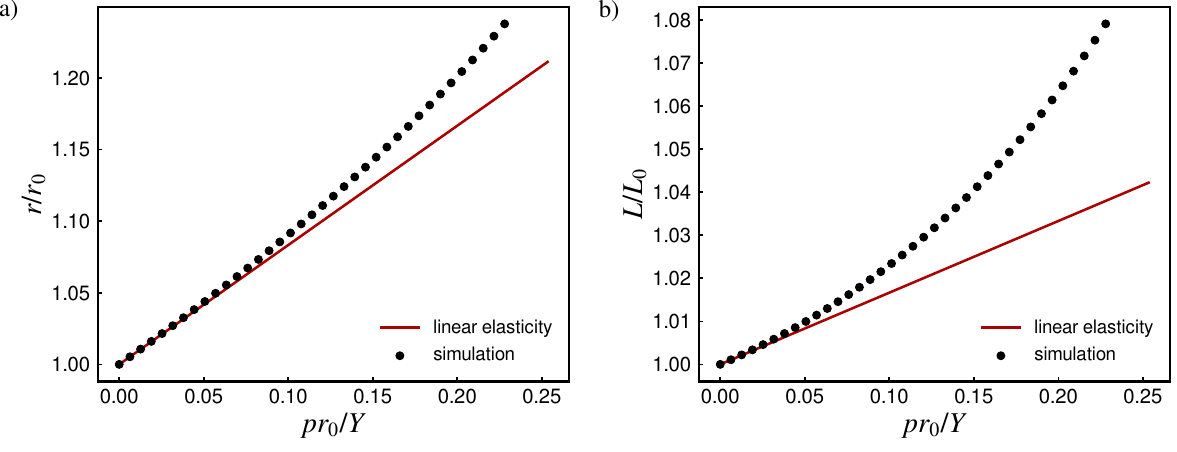}
    \caption[Non-linearity of the mesh]{Non-linearity of the mesh. a) Radius as a function of pressure. b) Length as a function of pressure. The red line is the expected response from linear elasticity.}
    \label{fig:non_linearity_mesh}
\end{figure}
\clearpage

\begin{figure}[t!]
    \centering
    \includegraphics[width=0.95\textwidth]{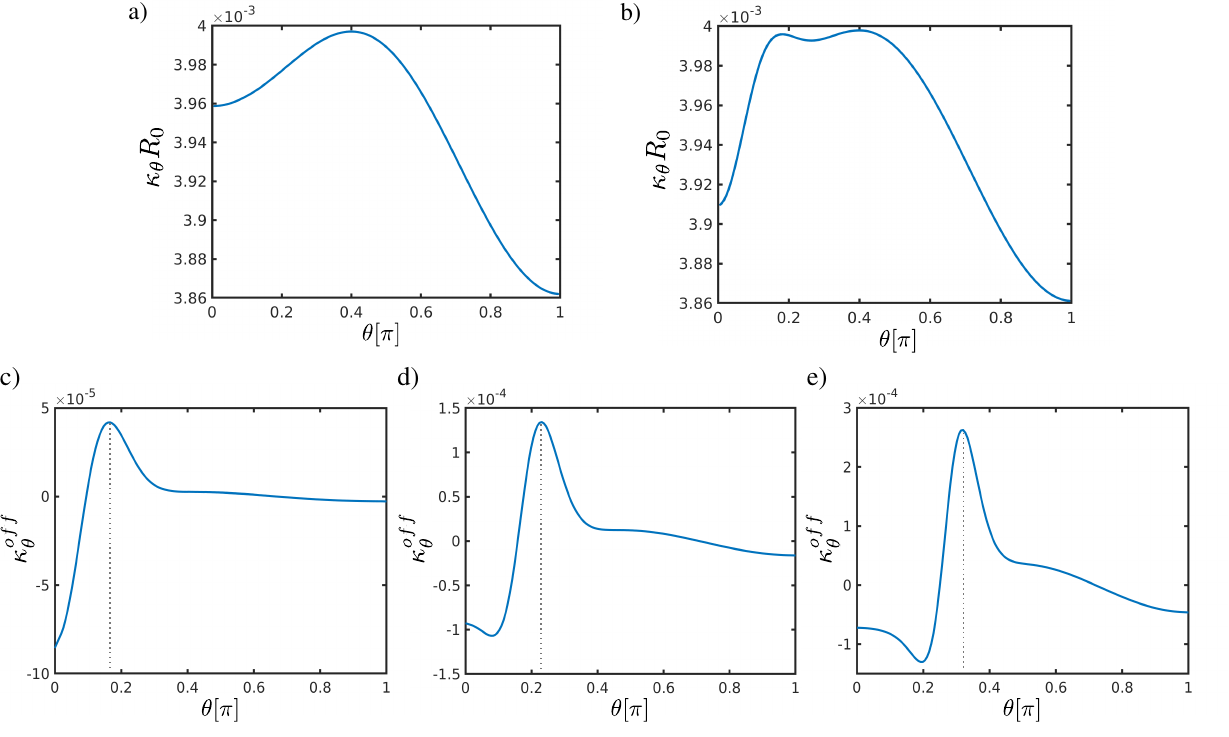}
    \caption[Determination of the extent of the wrinkles]{Determination of the extent of the wrinkles, $\theta_w$. a) Curvature along the circumferential direction $\kappa_\theta$ immediately before wrinkling ($\kappa/\kappa_w = 0.99$). b) $\kappa_\theta$ at the onset of wrinkling ($\kappa/\kappa_w = 1.00$). The extent of the wrinkles is obtained from the peak in $\kappa_\theta^{off}= \kappa_{\theta\theta}(\theta, \kappa) - \kappa_{\theta\theta}(\theta, \kappa_w)$. Examples for $\kappa/\kappa_w = $ 1.02, 1.13 and 1.4 are shown in panels c), d) and e) at $\hat{p} = 0.026$. The position of the peak, indicated with a vertical dashed line, serves as a signature for the extent of the wrinkles, $\theta_w$.}
    \label{fig:extent_wrinkles}
\end{figure}
\clearpage

\begin{figure}[t!]
    \centering
    \includegraphics[width=0.95\textwidth]{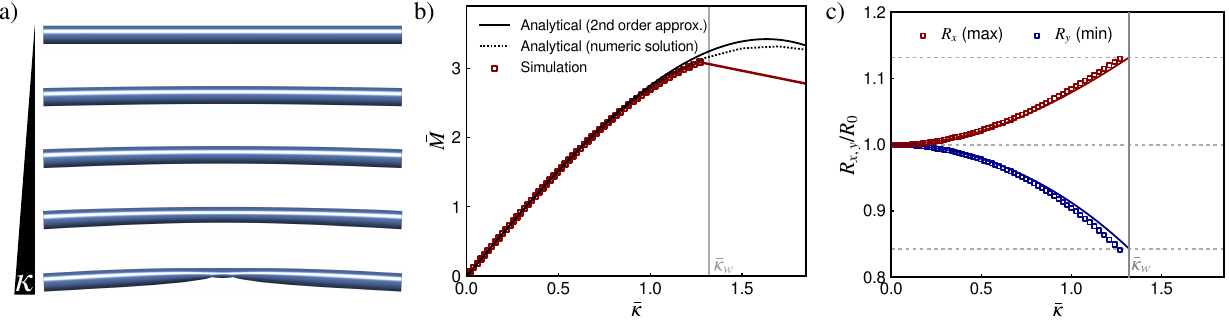}
    \caption[Mechanical response to bending in unpressurized tubes]{Mechanical response to bending in unpressurized tubes. a) Simulation results at progressively increasing induced curvatures. b) Torque-curvature response. The theoretical predictions considering the full expression and second order approximation (Equation \ref{eq:torque_shell}) are shown in dotted and continuous black lines, respectively. The expected buckling curvature is indicated by the vertical line ($\bar{\kappa}_w = 1.320$). c) Evolution of the semi-major and semi-minor axes of the cross-section of tube during bending. The dots are the simulation results. Continuous lines are the theoretical result determined from the numerical integration of Equations \ref{eq:rx_integral} and \ref{eq:ry_integral}. The horizontal dashed lines indicate the expected radii at the wrinkling curvature ($R_x =$ 1.1318; $R_y$ = 0.8430).}
    \label{fig:unpress_data}
\end{figure}
\clearpage

\begin{figure}[t!]
    \centering
    \includegraphics[width=0.99\textwidth]{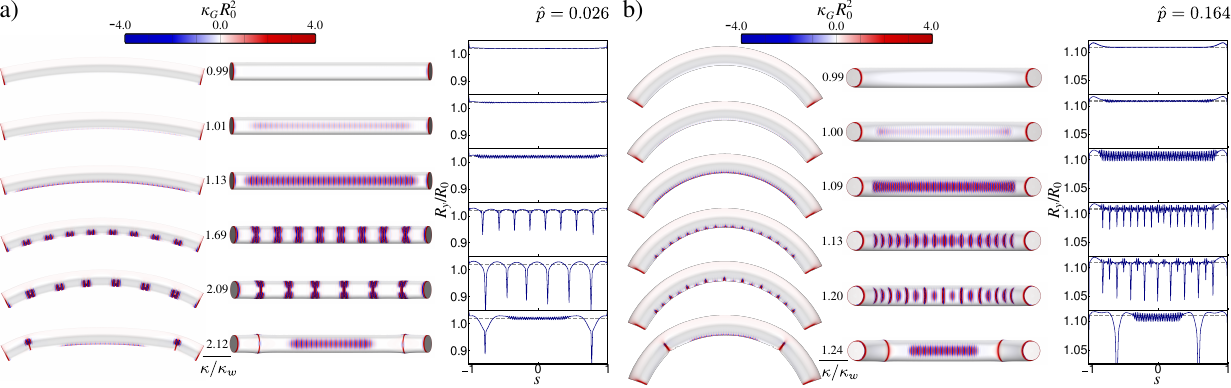}
    \caption[Wrinkling patterns during bending]{Wrinkling patterns during bending. a) Resulting configurations after the wrinkling transition for $\hat{p} \equiv pR/(Et) = 0.026$ at progressively increasing curvatures $\kappa/\kappa_w$ from top to bottom, where $\kappa_w$ is the curvature at the onset of wrinkling. The surface is colored according to Gaussian curvature, $\kappa_G$ \citep{meyers_diff_geo_mesh}.  Radius at the bottom of the tube $R_y$ along the arc length of the tube $s$ (normalized to the range -1 to 1). The pictures at $\kappa/\kappa_w = 1.69$ and $\kappa/\kappa_w = 2.09$  are representative examples of the system in the multi-wavelength state (MWS) in the low and high wavelength substates, respectively. 
    b) Configurations after the wrinkling transition for $\hat{p} = 0.164$ and $R_y$ as a function of $s$. The pictures at $\kappa/\kappa_w = 1.13$ and $\kappa/\kappa_w = 1.20$ are representative examples of the system in the MWS at the low and high wavelength substates, respectively.}
    \label{fig:wrinkles_general_show_comp_press}
\end{figure}
\clearpage

\begin{figure}[t!]
    \centering
    \includegraphics[width=0.75\textwidth]{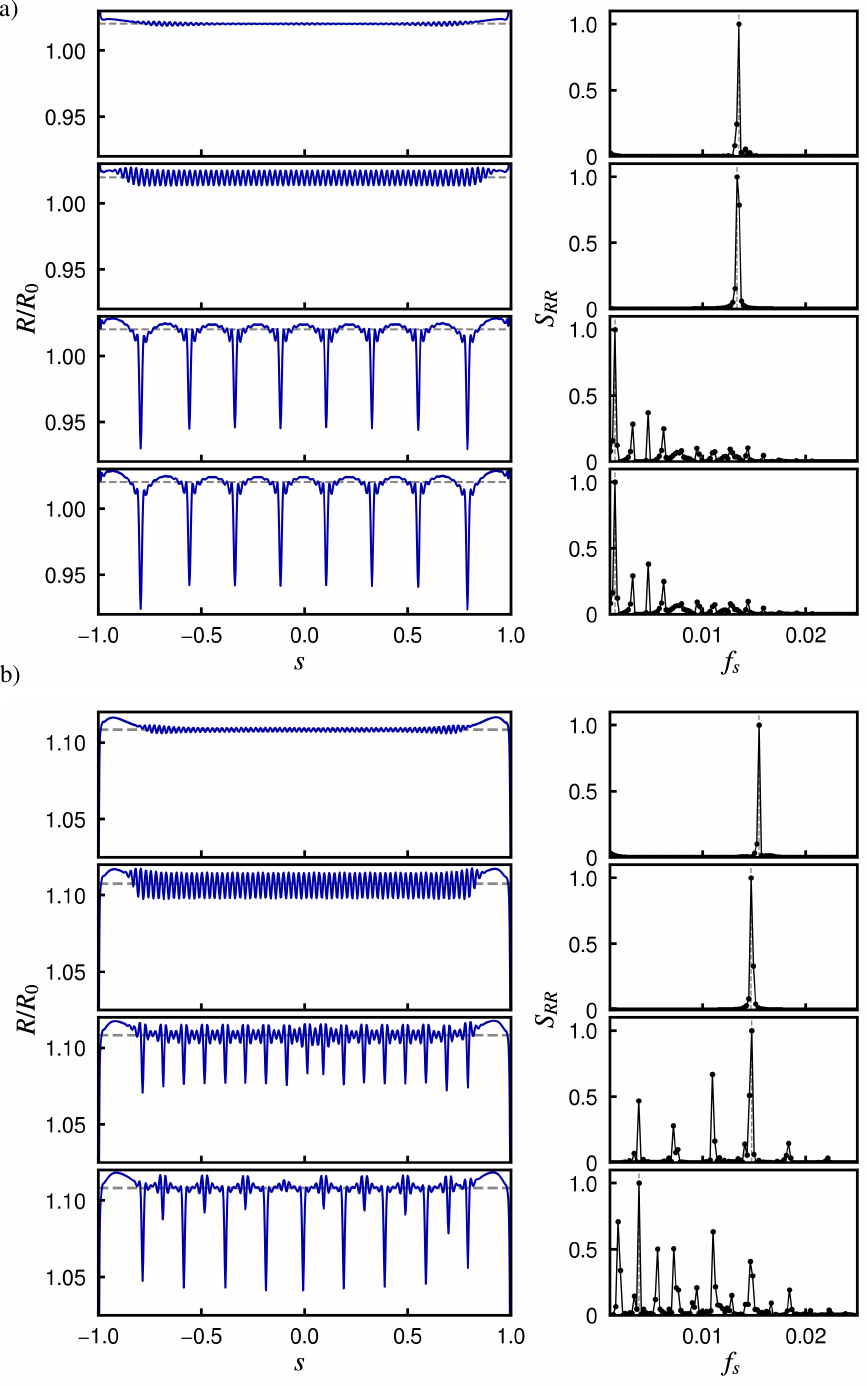}
    \caption[Influence of pressure on the wrinkle patterns]{Influence of pressure on the wrinkle patterns. a) Radii from the center of mass of the tube to the bottom side of the tube (\emph{left}). Power spectral density of the $R(s)$, \emph{right}. Data for $\hat{p} = $ 0.026. The curvatures are, from top to bottom, $\kappa/\kappa_w$ = 1.01, 1.13, 1.69, 2.09. b) Radii, \emph{left}, and power spectral density of the $R(s)$ function, \emph{right}. Data for $\hat{p} = 0.164$. The curvatures are, from top to bottom, $\kappa/\kappa_w$ = 1.00, 1.09, 1.13, 1.20. The power spectral density is defined as $S_{RR}(f_s) = |\tilde{R}(f_s)|^2$, with $\tilde{R} = \sum_{s=0}^L R(s) \exp(isf_s)$ the discrete Fourier transform of the radius. }
    \label{fig:amplitude_psd_wrinkles}
\end{figure}
\clearpage

\begin{figure}[t!]
    \centering
    \includegraphics[width=0.95\textwidth]{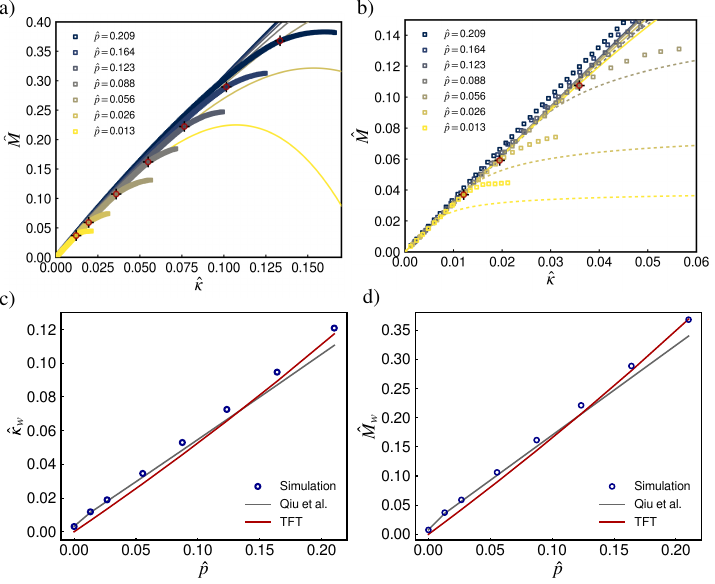}
    \caption[Comparison of analytical models for $\kappa$ and $M$]{Comparison analytical models for $\hat{\kappa}$ and $M$. a) Torque-curvature response. Model considering non-linear stress-strain relations. b)  Torque-curvature response. Detail at low pressures and comparison between tension field theory (dashed lines) and the model considering non-linear stress strains relations (continuous lines).  c) Curvature $\kappa_w$ as a function of pressure. Continuous lines are the predictions given by Equations \ref{eq:kappa_w_tft} and \ref{eq:kappa_w}. d) Torque $\hat{M}_w$ at the wrinkling transition as a function of pressure. Continuous lines are the predictions given by Equations \ref{eq:torque_tft_linear} and \ref{eq:torque_curv_qiu}. The rescaled medial axis curvature is related with the inner most curvature by $\hat{\mathcal{K}} = \hat{\kappa}/(1 + \hat{\kappa})$. }
    \label{fig:torque_kw_mw_comp}
\end{figure}
\clearpage

\begin{figure}[t!]
    \centering
    \includegraphics[width=0.95\textwidth]{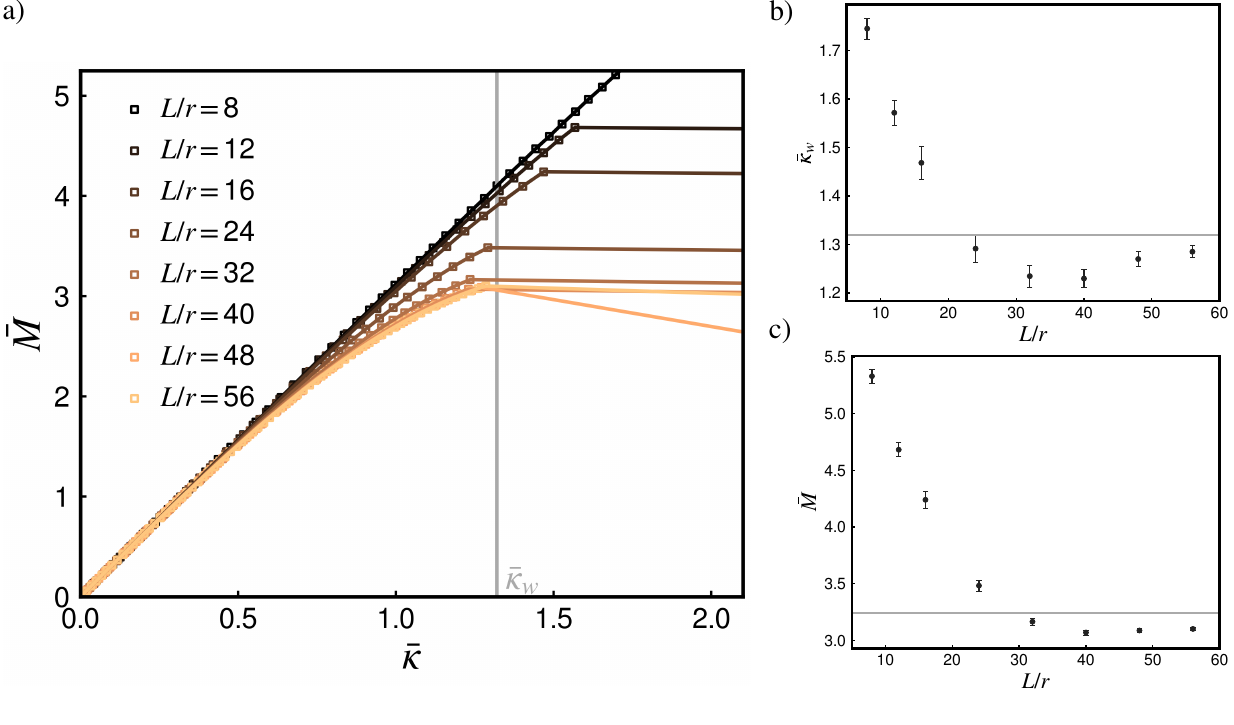}
    \caption[Impact of tube's length on mechanic parameters (non-pressurized)]{Impact of tube's length on mechanic parameters (non-pressurized). a) Torque-curvature curves for different lengths $L$ of the tube. b) Curvature at wrinkling $\bar{\kappa}_w$ depending on lengths. c) Torque at the onset of wrinkling $\bar{M}_w$ for different lengths. The continuous lines are the theoretical expectations (see unpressurized section in the main text).}
    \label{fig:mechanics_length_non_press}
\end{figure}
\clearpage

\begin{figure}[t!]
    \centering
    \includegraphics[width=0.95\textwidth]{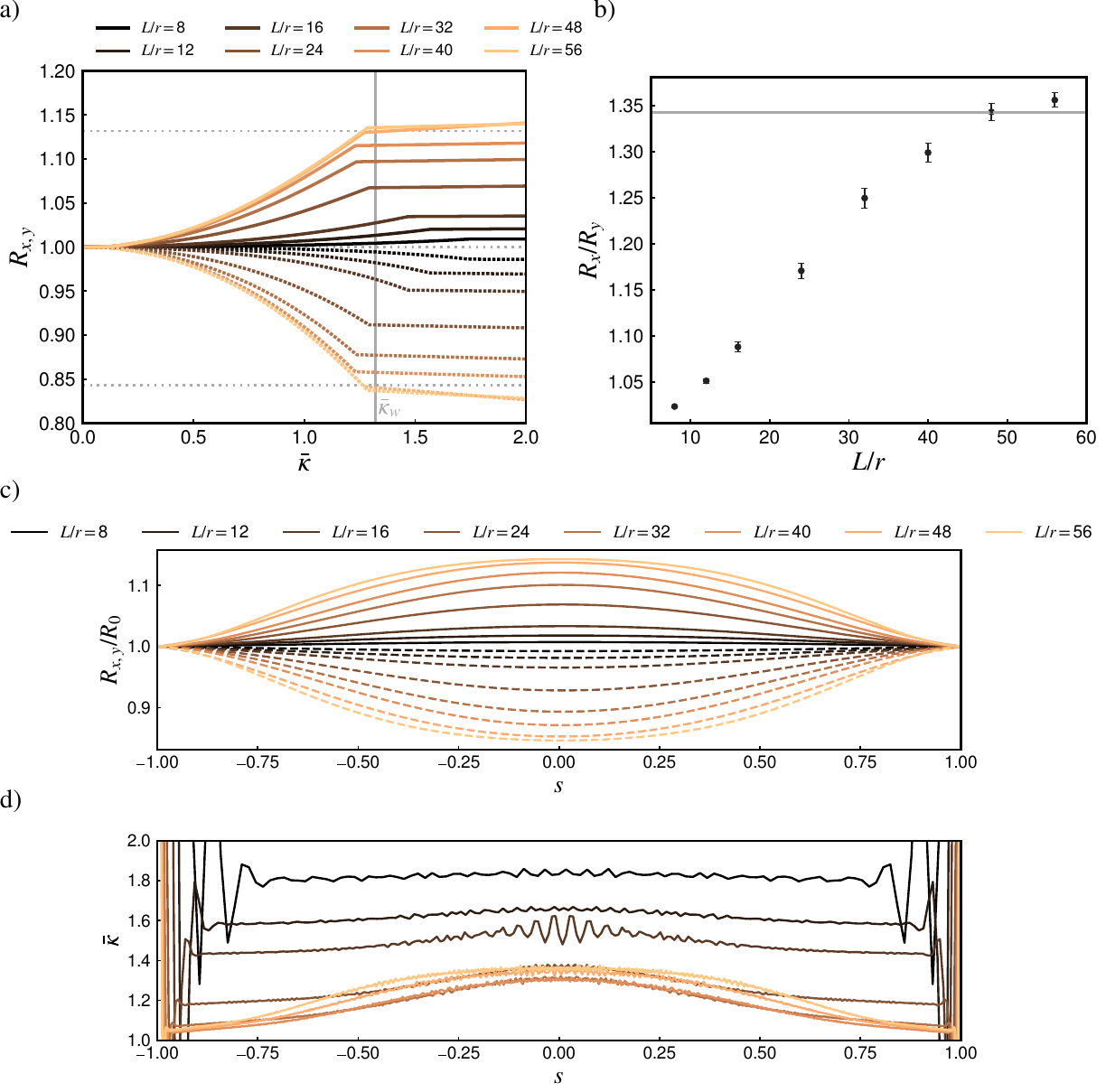}
    \caption[Impact of tube's length on geometric parameters (non-pressurized)]{Impact of tube's length on geometric parameters (non-pressurized). a) Maximum and minimum cross-section radii ($R_x$ and $R_y$) of the cross section as a function of the curvature for different lengths $L$ of the tube (ratio considering the undeformed shape). b) Ratio between $R_x$ and $R_y$ at the onset of wrinkling for different lengths.  Radii obtained in the 10\% central region. c) Dependency of the maximum and minimum radii $R_x$ and $R_y$ (dashed lines) as a function of $s$,  the arc length distance to the center of the tube, (scaled to the interval -1 to 1), for different lengths c) Curvature of the tube $\kappa$ as a function of $s$. Data in panels c) and d) taken from to immediate step before the wrinkling transition.}
    \label{fig:geometry_length_non_press}
\end{figure}
\clearpage

\begin{figure}[t!]
    \centering
    \includegraphics[width=0.95\textwidth]{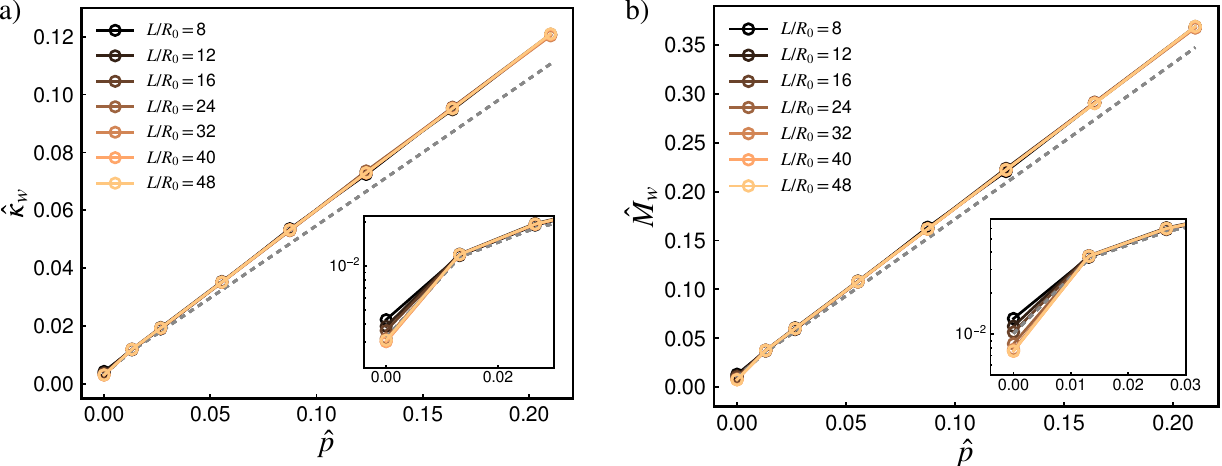}
    \caption[Impact of tube's length on  mechanic parameters (pressurized)]{Impact of tube's length on mechanic parameters (pressurized). a) Curvature at wrinkling $\bar{\kappa}_w$ depending on length. b) Torque at the onset of wrinkling $\bar{M}_w$ for different lengths. The dashed lines are the theoretical expectation (Equations \ref{eq:kappa_w} and \ref{eq:torque_curv_qiu}). The insets are a detail of the lowest pressures regime.}
    \label{fig:mechanics_length_press}
\end{figure}
\clearpage

\begin{figure}[t!]
    \centering
    \includegraphics[width=0.95\textwidth]{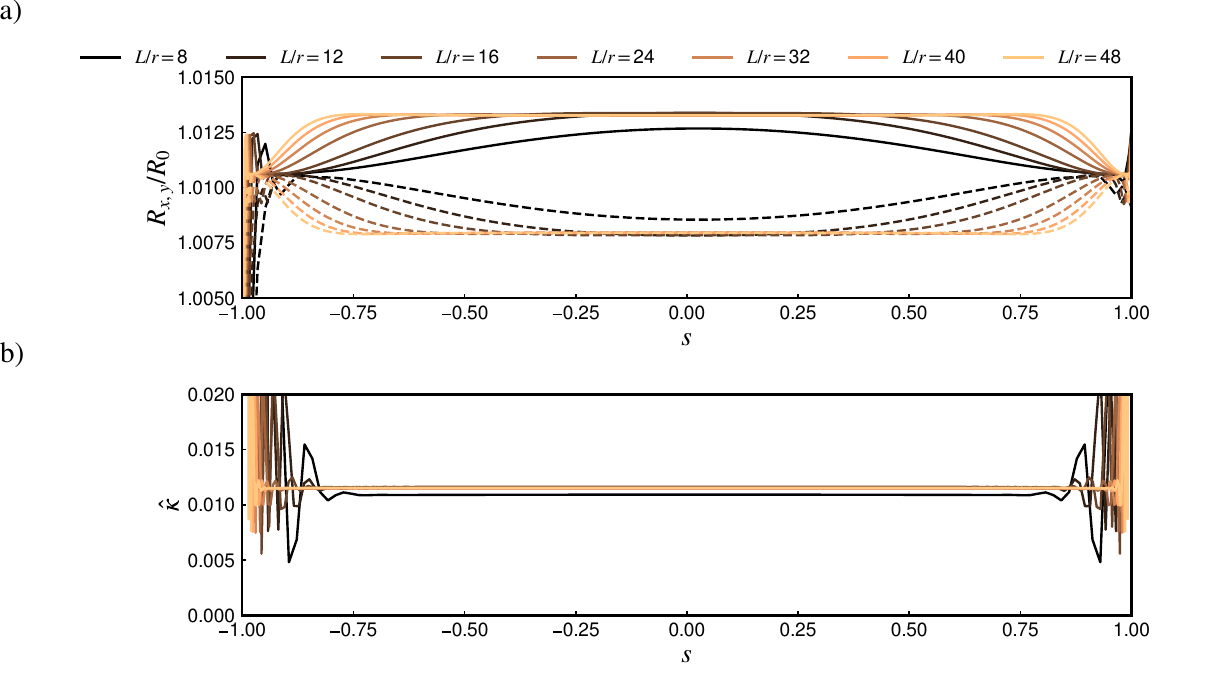}
    \caption[Impact of tube's length on geometric parameters (pressurized)]{Impact of tube's length on geometric parameters (pressurized). a) Dependency of the semimajor ($R_x$) and semiminor ($R_y$) radii immediately before the onset of wrinkling $\bar{\kappa}_w$ for different lengths $L$ depending on the position along the contour of the tube, $s$. b) Local curvature $\hat{\kappa}$ as a function of the contour length.}
    \label{fig:geometry_length_press}
\end{figure}
\clearpage

\begin{figure}[t!]
    \centering
    \includegraphics[width=0.83\textwidth]{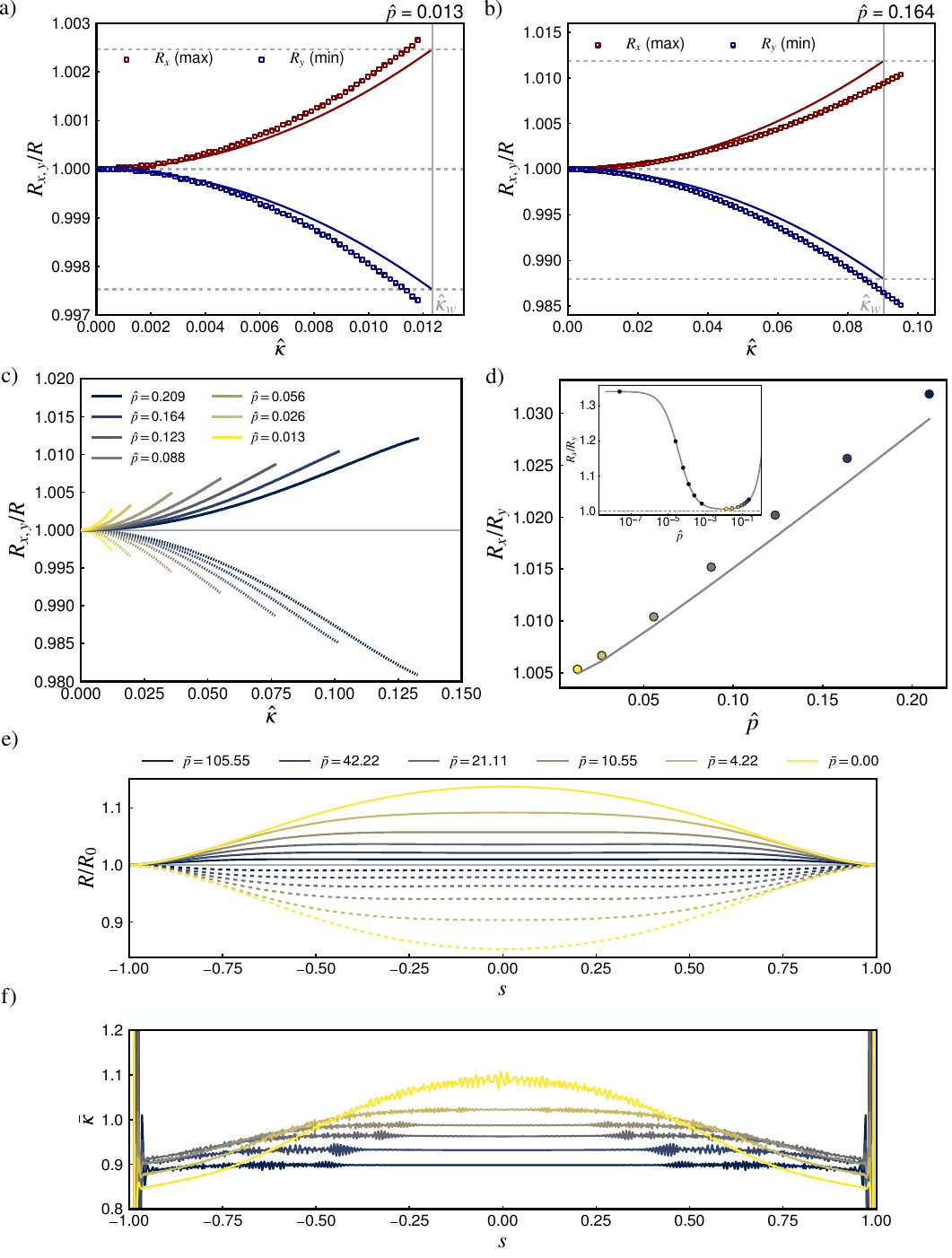}
    \caption[Geometry in pressurized tubes during bending]{Geometry of pressurized tubes during bending.
    a) and b) Evolution of the semi-major and semi-minor axes of the cross-section of tube
    during bending. The dots are simulation results. Continuous lines are the theoretical result determined from the numerical
    integration of Eqs. \ref{eq:rx_integral} and \ref{eq:ry_integral}. The horizontal dashed lines indicate the expected radii at the wrinkling curvature. Results for $\hat{p} = 0.013$ (a) and $\hat{p} = 0.164$ (b).
    c) Cross-section dimensions for the pressures shown in the main document. The deviation is $<$ 2\% for all the pressures (compare with Figure \ref{fig:geometry_length_non_press}a). Low pressures suppress the dependency on the radius and the curvature on the arc length. d) Eccentricity at the onset of wrinkling as a function of pressure for the pressures  studied on the main document. The grey line is corresponds to the theoretical predictions obtained from Equation \ref{eq:rxry}. The inset shows the theoretical prediction over a wide range of pressures, highlighting the non-monotonic response of the deformations with pressure. The pressures studied in the main documents are shown with coloured dots. Very low pressures are shown with black dots. 
    e) Deformations of the cross-section as a function of the normalized arc length of the tube, $s$ at the onset of wrinkling at low pressures (corresponding to the black dots in panel d). The continuous lines is the semi-major radius $R_x$ and the dotted line is the semi-minor radius $R_y$.
    f) Curvature as a function of $s$ at the onset of wrinkling at low pressures. The pressure in the \emph{shell} normalization is defined as $\bar{p} := pR^2$. The relation with the \emph{balloon} normalization is $\hat{p} = \alpha^2\bar{p}$. Here $\hat{p}$ are on the order  $10^{-5}$ - $10^{-4}$. Data for $L/R_0 = $ 48.}
    \label{fig:cross_section_radii_press}
\end{figure}
\clearpage

\begin{figure}[t!]
    \centering
    \includegraphics[width=0.65\textwidth]{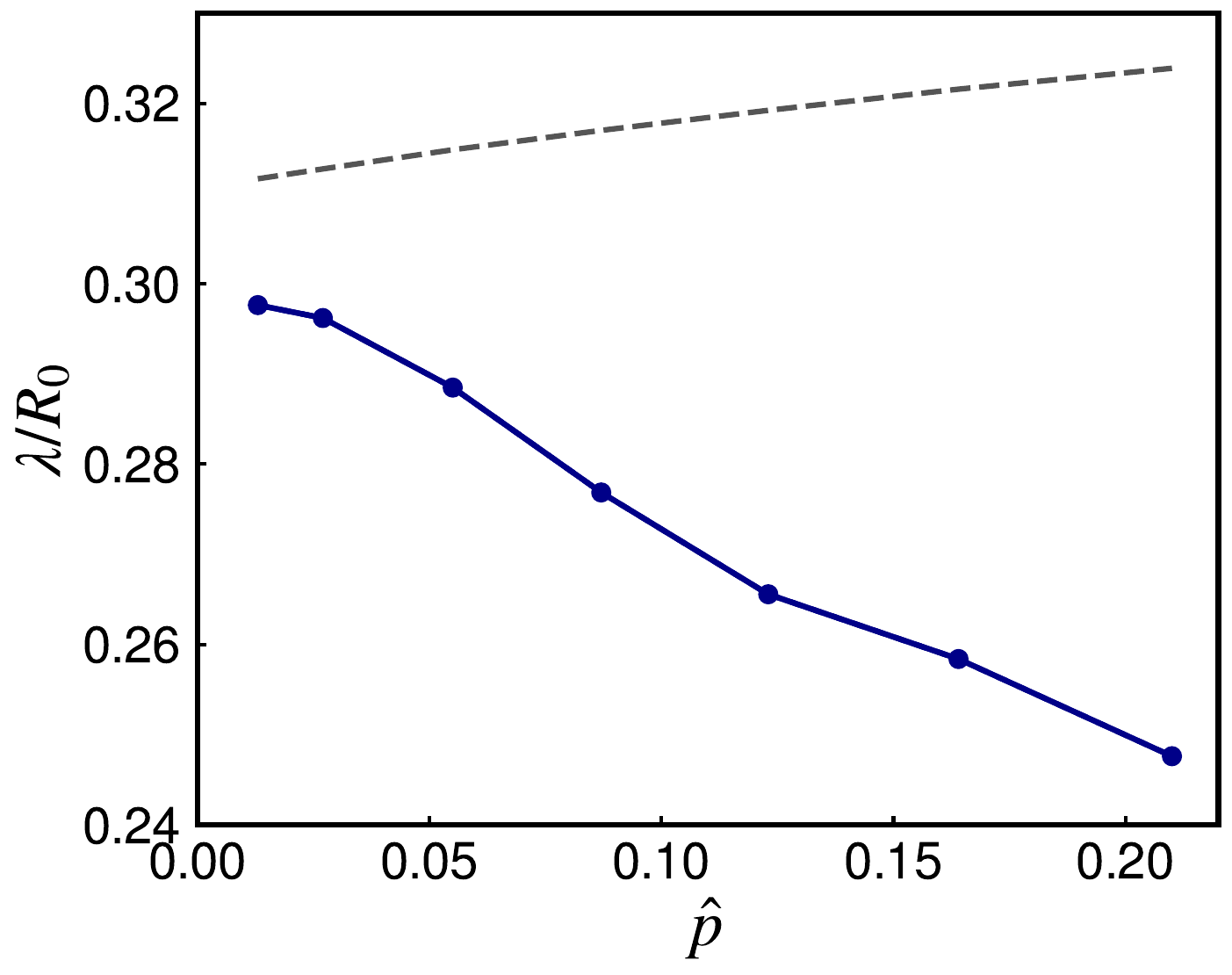}
    \caption[Dependency of the wavelength $\lambda$ of the sinusoidal wrinkles]{Dependency of the wavelength $\lambda$ of the sinusoidal wrinkles. Values determined at the onset of wrinkling. The dashed line is the predicted $\lambda_\mathrm{sin}$ of an unpressurized bent tube at the onset of wrinkling ~\citep{seide_weigarten_wavelength_bent_tube}: $\lambda = 2\pi \left(\frac{B}{Y}R_{\theta}^2\right)^{1/4}$. This simple formula predicts an implicit increase of $\lambda$ with pressure as a result of the expansion of $R_\theta$ . In contrast, we observe that pressurization leads to a decrease in $\lambda$. }
    \label{fig:lambda_press}
\end{figure}
\clearpage

\begin{figure}[t!]
    \centering
    \includegraphics[width=0.65\textwidth]{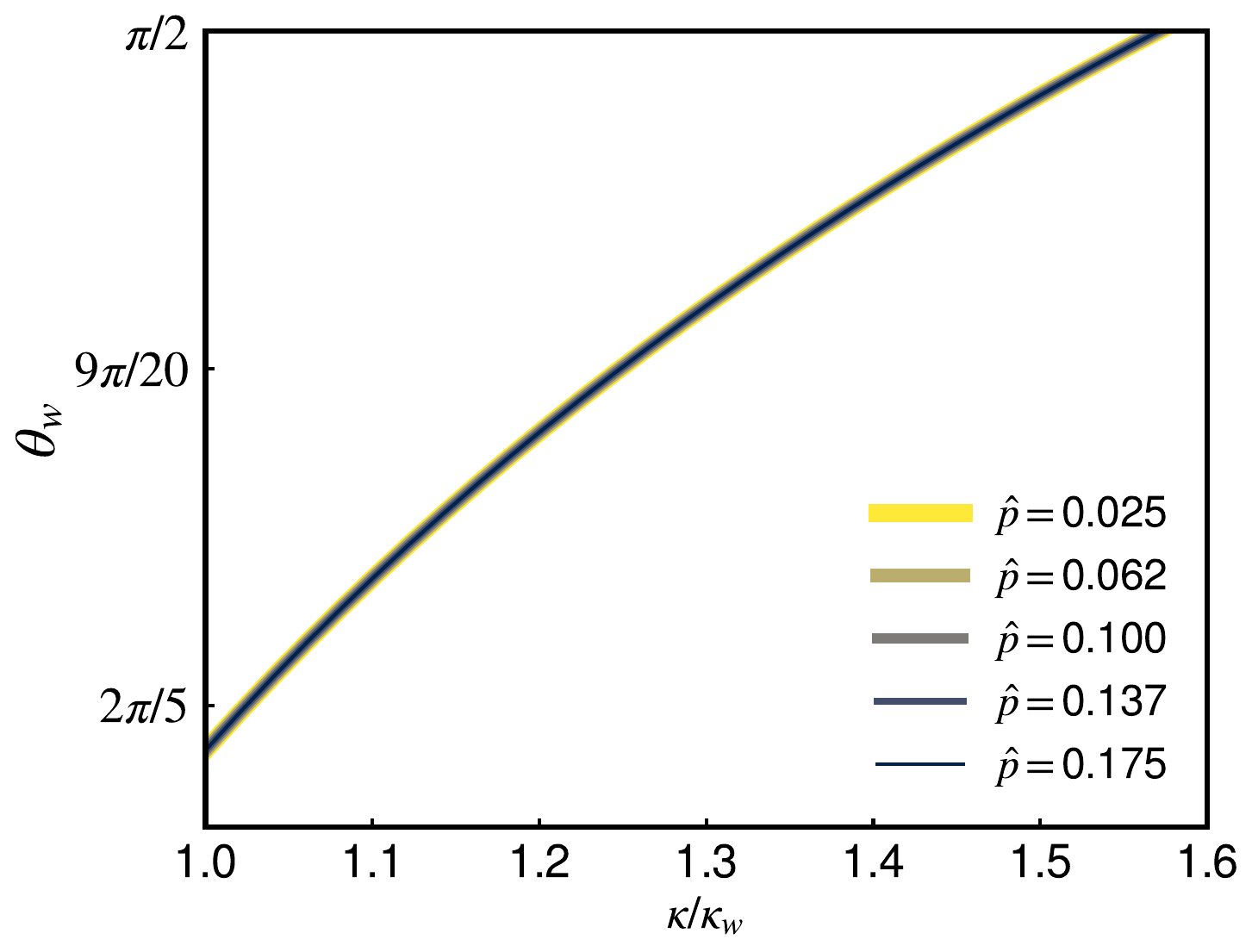}
    \caption[Expansion of wrinkles $\theta_w$ (tension field theory)]{Expansion of wrinkles $\theta_w$ from tension field theory as a function of curvature $\kappa_w$ for different pressures (Equation \ref{eq:force_balance}). The area expanded by the wrinkles is defined as $\theta_w := \pi - \theta_0$.}
    \label{fig:tft_wrinkle_expansion}
\end{figure}
\clearpage

\begin{figure}[t!]
    \centering
    \includegraphics[width=0.65\textwidth]{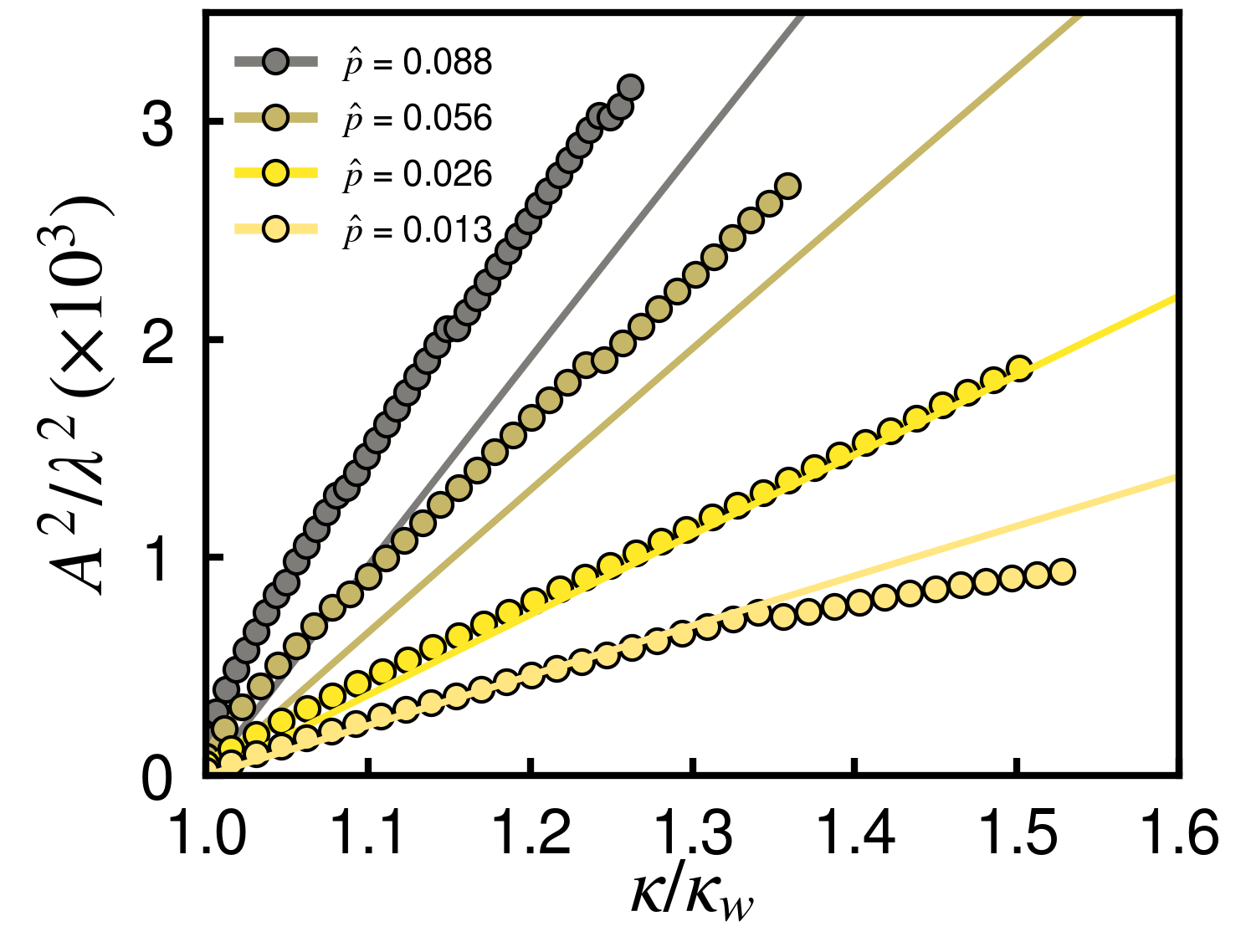}
    \caption[$(A/\lambda)^2$ in the sinusoidal regime during bending]{$(A/\lambda)^2$ in the sinusoidal regime as a function of curvature. The continuous lines are the predictions from Eq.~\ref{eq:amp_lambda_sq}.}
    \label{fig:amp_lampbda_sq}
\end{figure}
\clearpage

\clearpage
\newpage

\bibliographystyle{apsrev4-1} 
\bibliography{sm}